# Unveiling the nature of cathodoluminescence from photon statistics

Sotatsu Yanagimoto[1], Naoki Yamamoto[1], Tatsuro Yuge[2,3], Takumi Sannomiya[1*], Keiichirou Akiba[1, 4*]

**Affiliations:**
[1] Department of Materials Science and Engineering, School of Materials and Chemical Technology, Institute of Science Tokyo, 4259 Nagatsuta, Midoriku, Yokohama, 226-8503, Japan

[2] Department of Physics, Shizuoka University, 836 Ohya, Surugaku, Shizuoka, 422-8529, Japan

[3] Graduate School of Science and Technology, Shizuoka University, 836 Ohya, Surugaku, Shizuoka, 422-8529, Japan

[4] Takasaki Institute for Advanced Quantum Science, National Institutes for Quantum Science and Technology (QST), 1233 Watanuki, Takasaki, Gunma, 370-1292, Japan

*Corresponding authors email:
sannomiya.t.aa@m.titech.ac.jp (T.S.) ; akiba.keiichiro@qst.go.jp (K.A.)

## Abstract

Cathodoluminescence (CL), the emission of light induced by accelerated free electrons, has been extensively utilized in various applications, such as displays, streak cameras, and high-spatial-resolution analysis of optical material, surpassing the diffraction limit of light. Despite its long history, the photon statistics of CL have only recently been examined, revealing unexpectedly large bunching of photons. Here we find that this peculiar photon bunching contains information of intervening excitation processes before the photon emission, which can be extracted from the photon statistics within each excitation event by a single free electron. Using this approach, we experimentally unveiled the statistical differences of coherent CL involving a single electromagnetic interaction process and incoherent CL involving multiple excitation processes. The developed formulation is universally applicable for particle generation processes in general to investigate the nature of cascade reactions.

## Introduction

Cathodoluminescence (CL), light emission by free electron excitation, has long been known since the discovery of electrons by Thomson in the late 19th century and utilized for cathode ray tube (CRT) displays for almost a century until recently. Nowadays, the capability of fast CL signal switching by electron beam (e-beam) is applied to streak cameras to visualize ultrafast phenomena. Another significant achievement of CL is to analyze the optical properties of materials using electron microscopes, offering much higher spatial resolution than purely optical means beyond the diffraction limit of light [1–3]. The CL process is typically categorized into two types, namely coherent CL, where the photons are emitted through pure electromagnetic interaction with the fast electron, and incoherent CL, which involves relaxation processes through the excitation of particles or quasi-particles, such as secondary electrons, plasmon or excitons [4]. While incoherent CL imaging and spectroscopy have been employed to analyze semiconductors and minerals since the 1970s [5–9], coherent CL has been more popularly used in the last decades in the field of nanophotonics and plasmonics, which allows visualizing the electromagnetic field at nanoscales [10–12]. For coherent CL, use of entanglement between free electrons and emitted photons have been recently proposed for quantum applications [13–16].

Regardless of its long history and prevalence, the nature of the free-electron excited light in terms of photon statistics has been examined only recently [17–19]. The CL photons show unusual statistics giving colossal bunching in the second order correlation ($g^{(2)}$) larger than two for almost all materials, including both coherent and incoherent CL even at room temperature [20–24]. This peculiar nature has been empirically described by semi-classical as well as quantum approaches, and it was elucidated that the CL photon statistics basically originate from random excitation by individual electrons [21,25,26]. Therefore, exclusion of the influence of excitation modulation by free electrons in the photon statistics, which is the indicator of the photon state, is essential to understand the photon generation processes; for instance, incoherent CL photon bunching is influenced by the excitation of bulk plasmons or secondary electrons[21,24].

In this study, we propose an approach to exclude the apparent electron excitation modulation effect and to extract the intrinsic CL photon statistics, which directly reflect the photon state within the single electron excitation event. This approach enables quantitative comparison between CL photon statistics $g^{(2)}_{\mathrm{CL}}$ and PL photon statistics $g^{(2)}_{\mathrm{PL}}$ or theoretical models, providing insights into the excitation and emission processes in materials. We experimentally quantify the CL photon statistics and verify the photon generation processes both for coherent CL with Poisson distributions and for incoherent CL with super-Poisson distributions. And finally, instead of the numerical approaches[27], we provide a general description of this statistical analysis including multiple cascade processes derived by a fully analytical approach.

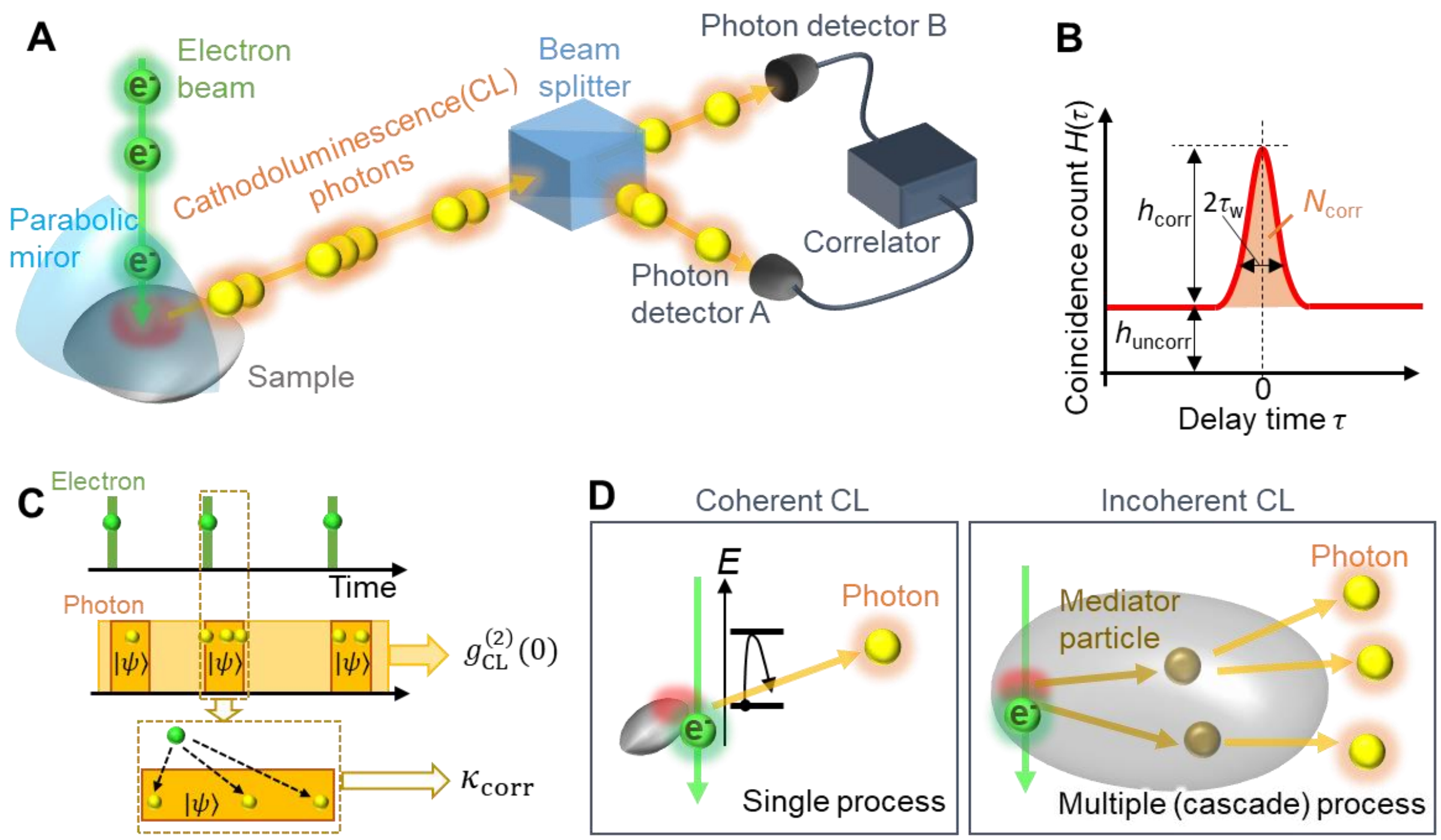

**Fig. 1. Photon statistics analysis of cathodoluminescence.**
(**A**) Schematic illustration of the photon statistics measurement in cathodoluminescence (CL). CL photons generated upon free electron excitation are introduced to a 50-50 beam splitter and counted by detectors A and B. (**B**) The coincidence events of the two detectors are counted as a function of the delay time difference between the two detectors to produce a coincidence histogram. CL typically shows a bunching feature with higher counts at the zero time delay. (**C**) Schematic diagram showing the relations of the excitation events by free electrons and the generated photon state. (**D**) Illustrations of the coherent and incoherent CL. In the coherent CL, photons are electromagnetically generated as a single process, keeping the electromagnetic coherence between the incident electron and generated photon, while in the incoherent CL, relaxation processes, including the generation of mediator particles are involved, through which the coherence between the incident electron and the generated photon is lost.

## Results and Discussion

### Extraction of CL photon statistics within single electron excitation event

We used a scanning transmission electron microscope (STEM) equipped with a Hanbury Brown-

Twiss (HBT) setup consisting of a 50:50 beam splitter, two photon counters and a correlator, as shown in Fig. 1A. The CL-HBT approach enables measuring the second order correlation function $g^{(2)}$ of the generated CL photons [27,28], which is obtained after normalizing the coincidence histogram by the flat uncorrelated background signal $h_{\mathrm{uncorr}}$ (Fig. 1B). $g^{(2)}$ of CL, which we now call $g_{\mathrm{CL}}^{(2)}$, typically shows peculiar bunching because the incident electrons randomly excite photons, resulting in sparsely distributed photon states with random vacuum intervals (Fig. 1C) [25]. When we consider a model where $n$ photons are excited with a probability $p_{\mathrm{n}}$ during a single electron injection event, the correlated counts $N_{\mathrm{corr}}$ (see Fig. 1B) and background $h_{\mathrm{uncorr}}$ are numbered as follows:

$$N_{\mathrm{corr}} = \frac{1}{4}\eta_{\mathrm{A}}\eta_{\mathrm{B}}T\frac{I_{\mathrm{e}}}{e}\sum_{n=0}^{\infty} p_{\mathrm{n}} n(n-1)$$

$$h_{\mathrm{uncorr}} = \frac{1}{4}\eta_{\mathrm{A}}\eta_{\mathrm{B}}T t_{\mathrm{bin}}\left(\frac{I_{\mathrm{e}}}{e}\right)^2\left(\sum_{n=0}^{\infty} p_{\mathrm{n}}\, n\right)^2 \tag{1}$$

where $\eta_{\mathrm{A(B)}}$ is the detection efficiency of the system for detector A (B), $t_{\mathrm{bin}}$ the time bin width, $I_{\mathrm{e}}$ the electron beam current, $e$ the elementary charge, and $T$ the integration time of the coincidence histogram (Detailed derivation process is described in Methods section). The total coincident count $N_{\mathrm{corr}}$ is spread within the typical correlation time width of $\tau_{\mathrm{w}}$, reflecting the lifetime of the photon source and the instrument response time. The correlation height $h_{\mathrm{corr}}$ indicated in Fig. 1B is defined as

$$h_{\mathrm{corr}} = \frac{N_{\mathrm{corr}} t_{\mathrm{bin}}}{2\tau_{\mathrm{w}}}. \tag{2}$$

Using Eq. 1 and 2, one can extract the correlation function at time delay zero as

$$g_{\mathrm{CL}}^{(2)}(0) = 1 + \frac{h_{\mathrm{corr}}}{h_{\mathrm{uncorr}}} = 1 + \frac{e}{2I_{\mathrm{e}}\tau_{\mathrm{w}}}\kappa_{\mathrm{corr}}. \tag{3}$$

Here, we introduced a correlation factor $\kappa_{\mathrm{corr}}$, which corresponds to the second order coherence of the emitted light within the single electron excitation event:

$$\kappa_{\mathrm{corr}} = \frac{\sum_{n=0}^{\infty} p_{\mathrm{n}} n(n-1)}{\left(\sum_{n=0}^{\infty} p_{\mathrm{n}}\, n\right)^2} = \frac{\langle n^2\rangle - \langle n\rangle}{\langle n\rangle^2}. \tag{4}$$

$\langle\cdots\rangle$ represents an ensemble average within a single electron excitation event. The expression of Eq. 3 clearly indicates that the experimentally obtained $g_{\mathrm{CL}}^{(2)}(0)$ includes the modulation factor $\frac{e}{2I_{\mathrm{e}}\tau_{\mathrm{w}}}$, as reported in previous work [21,27], which disguises the photon statistics of emission excited by individual electrons. Thus, by calculating $\kappa_{\mathrm{corr}}$ from $g_{\mathrm{CL}}^{(2)}(0)$ according to Eq. 3, one can extract the inherent statistics of generated photons within the single excitation event (i.e., single electron incidence) without the excitation modulation effect by free electrons [25]. In other words, $\kappa_{\mathrm{corr}}$ corresponds to $g^{(2)}(0)$ of $|\psi\rangle$, removing the influence of vacuum (Fig. 1C). From Eq. 4, it is evident that the Poisson distributions of photons (such as coherent state) give $\kappa_{\mathrm{corr}} = 1$.

**Coherent CL**

We start evaluating the correlation factor $\kappa_{\mathrm{corr}}$ for the coherent CL. In the coherent CL, the photons are directly generated by free electrons only through an electromagnetic interaction [4]. Thus, coherent CL involves only a single process (Fig. 1C). For such coherent CL, the photon state by each electron excitation event can reasonably be assumed as a coherent state [29]. Then, the photon statistics should follow the Poisson distribution, and $\kappa_{\mathrm{corr}} = 1$ is expected. (Even if electron-photon entanglement arises in the coherent CL process, the photon statistics are the same as the coherent state.) We verify these coherent CL statistics using various systems, as summarized in Fig. 2.

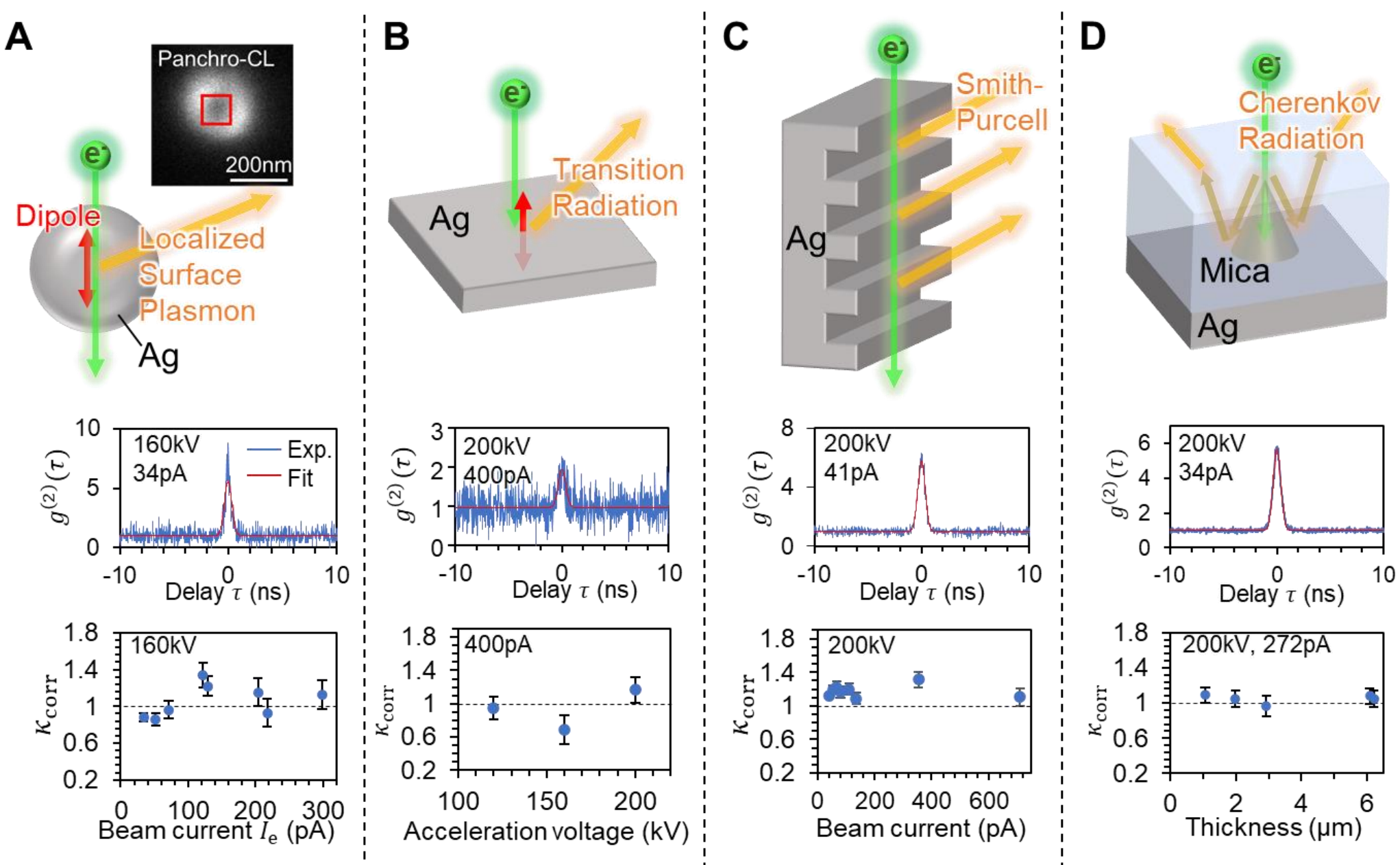

**Fig. 2. Statistical analysis results of coherent cathodoluminescence.**
(**A**) Localized surface plasmon emission, (**B**) transition radiation, (**C**) Smith-Purcell radiation, and (**D**) Cherenkov radiation. The second row in each panel shows the second order correlation function of CL photons in a representative condition. The correlation factors extracted for various conditions are plotted in the third row.

First, we evaluate the correlation factor $\kappa_{\mathrm{corr}}$ of a single dipole radiation, which is the simplest example of coherent CL. We use a silver sphere with a diameter of 220 nm as a coherent CL source, which supports localized surface plasmons (LSPs), i.e. collective oscillations of free electrons of a metal particle, and can be treated as a single electric dipole, as shown in Fig. 2A [30]. The second row of Fig. 2A shows the correlation function, presenting the typical bunching feature. The obtained correlation time $\tau_{\mathrm{w}}$ (Fig. 1B) is approximately 400 to 500 ps, which is limited by the temporal resolution of the instrument (see supplementary information Sec.), showing that the lifetime of coherent CL is extremely short and cannot be resolved [31,32]. The height of the bunching peak is inversely proportional to the e-beam current, meaning that the bunching is caused by the excitation modulation (see Eq. 3). The correlation factor $\kappa_{\mathrm{corr}}$ is independent of the e-beam current and shows approximately constant values $\kappa_{\mathrm{corr}} \sim 1$. This verifies that the single process of photon generation in a coherent LSP dipole follows the Poisson statistics.

As the second example of a simple coherent dipole case, we examine transition radiation (TR). TR can be considered as an effective dipole generated at the interface between metal and dielectric, as illustrated in Fig. 2B [4]. We utilize a 200 nm silver film and evaluate the acceleration voltage dependency of $\kappa_{\mathrm{corr}}$. The $g^{(2)}_{\mathrm{CL}}(\tau)$ curve shows a characteristic fast decay similar to the LSP case. We changed the acceleration voltage in order to manipulate the expected value of the photon number generated by a single electron incidence since TR excitation probability depends on the acceleration voltage[4]. The obtained correlation factor $\kappa_{\mathrm{corr}}$ values in Fig. 2B are independent of the acceleration voltage and remain $\sim 1$, which is basically identical to the LSP case. We here note that even when a free electron produces TR and another coherent emission excited through SPPs from a plasmonic structure at a distance in a combined structure, the $\kappa_{\mathrm{corr}}$ value still stays around one ($\kappa_{\mathrm{corr}}$ = 0.94 ± 0.09) (see Supplementary Information). This fact indicates that the distribution of the target electromagnetic wavefunction does not play a prominent role in the generated photon statistics as long as the CL process is coherent.

We also test "thicker" structures compared to the simple dipole cases presented above. The first example of the thick structure is the Smith-Purcell (S-P) radiation, another type of coherent CL, which is produced when a charged particle travels along a periodic structure (Fig. 2C) [33,34]. Thanks to the long interaction length along the beam path, S-P radiation attains strong intensities even without electrons directly hitting the structures, i.e., aloof excitation. We use a silver-coated periodic structure (periodicity: 800nm, lattice height: 350nm). The correlation factor $\kappa_{\mathrm{corr}}$ is acquired in the aloof condition with the e-beam located 100 nm away from the structure surface. The value of $\kappa_{\text{corr}}$ in this configuration is approximately one, independent of the e-beam current, which is again consistent with the Poisson statistics of the coherent CL. We noticed that the $\kappa_{\text{corr}}$ value exceeds one ($\kappa_{\text{corr}} > 10$) when the e-beam hits the structure directly, which is investigated by line-scanning the e-beam from the specimen to vacuum (see Supplementary Information). We attribute this effect to the reduction of the effective beam current, or underestimation of the modulation factor, due to the e-beam scattering or absorption since the smaller beam current shows larger apparent bunching, as shown in Eq. 3 [25,26] (see Supplementary Information).

As the last example of coherent CL with a long interaction length, Cherenkov radiation (CR) is examined. We use a natural mica membrane with a refractive index of ~1.6, which gives the threshold electron energy of ~140 keV for CR (see Supplementary Information). Since CR is emitted in the forward direction of the electron's travel (Fig. 2D), we deposited a silver layer on the bottom side of the exfoliated mica membrane, which reflects the CR upwards, in order to efficiently collect CR [35]. Although we observe the inclusion of the impurity (incoherent) emission in the spectrum for lower acceleration, at the acceleration voltage of 200 kV we observe only dominant CR [20,35](see Supplementary Information). For this sample, we changed the sample thickness to manipulate the expected value of the photon number per electron since the excitation probability is proportional to the interaction length, i.e. thickness. The obtained $\kappa_{\text{corr}}$ values are around one independent of the thickness of the sample at 200 kV acceleration, as shown in Fig. 2D. Therefore, the thickness-independent trend shown in Fig. 2D, similar to the discussion on TR with varying acceleration voltage, shows that the correlation factor reflects only the statistical variance in the photon number distribution, but not the excitation probability. At the lower acceleration voltages, we observed lower $\kappa_{\text{corr}}$ values ($0.48 \pm 0.01$ and $0.26 \pm 0.07$ for 160 and 120 kV, respectively) (see Supplementary Information). Because the lifetime of the impurity emission, which becomes non-negligible at lower acceleration, is much longer than CR, the inclusion of this incoherent impurity CL significantly contributes to the uncorrelated background ($h_{\text{uncorr}}$ in Fig. 1B) and decreases the apparent $\kappa_{\text{corr}}$ value.

As explained above for all the coherent CL systems, we successfully excluded the effect of electron modulation and extracted the inherent photon statistics within a single excitation event i.e., $\kappa_{\text{corr}}$, which follows the theoretically expected Poisson distribution.

## **Incoherent CL**

In incoherent CL, the photons are generated through multi-step (cascade) processes (Fig. 1D). Although the photon statistics in incoherent CL are previously analyzed [21,26], we here develop an analytical description of incoherent CL by considering the particle number probability of each excitation step. The first step is the excitation event by free electron.

As described in Fig. 1D, photons are excited through mediator particles, such as bulk plasmons or secondary electrons, thus involving relaxation processes [4]. As the simplest model, we can consider two cascade excitation processes: (1) fast electrons generate mediator particles, and (2) the mediator particles generate photons. The correlation factor for this incoherent CL process is then expressed as follows:

$$\kappa_{\mathrm{corr}} = \kappa_{\mathrm{corr}}^{\mathrm{med}} + \frac{1}{\langle n^{\mathrm{med}} \rangle} \kappa_{\mathrm{corr}}^{\mathrm{ph}} \,. \tag{5}$$

Here, we introduced the correlation factor for the mediator particle $\kappa_{\mathrm{corr}}^{\mathrm{med}} = \frac{\left\langle (n^{\mathrm{med}})^2 \right\rangle - \langle n^{\mathrm{med}} \rangle}{\langle n^{\mathrm{med}} \rangle^2}$ and that of the photon generation from the mediator particle $\kappa_{\mathrm{corr}}^{\mathrm{ph}} = \frac{\left\langle (n^{\mathrm{ph}})^2 \right\rangle_{1\mathrm{med}} - \langle n^{\mathrm{ph}} \rangle_{1\mathrm{med}}}{\langle n^{\mathrm{ph}} \rangle_{1\mathrm{med}}^2}$, where the suffix “1med” for the angle bracket denotes averaging per mediator particle (see Supplementary Information for the detailed derivation). When each single process can be considered to follow the Poisson statistics, $\kappa_{\mathrm{corr}}^{\mathrm{med}} = \kappa_{\mathrm{corr}}^{\mathrm{ph}} = 1$ holds, which results in a similar expression as the previous study [21]. Then, $\langle n^{\mathrm{med}} \rangle$ can be calculated from the experimentally obtained $\kappa_{\mathrm{corr}}$. According to Eqs. 3 and 5, the “excitation efficiency" in the previous studies assuming bulk plasmons[21,36] corresponds to $\langle n^{\mathrm{med}} \rangle$ in the limit of low excitation. We note that Eq. 5 can also be derived from a master equation approach assuming a cascade-type particle generation reaction as shown in Supplementary Information.

We apply this analysis to incoherent CL of $Y_2SiO_5$: Ce (YSO), as shown in Fig. 3A. We confirmed that $\kappa_{\mathrm{corr}}$ for YSO remains constant with respect to the electron beam current, indicating that photon statistics excited by individual electrons are extracted (see Supplementary Information). Since the average number of mediator particles per electron $\langle n^{\mathrm{med}} \rangle$ should depend on the thickness, one can vary $\langle n^{\mathrm{med}} \rangle$ value by sample thickness. We scanned the e-beam from the thin position A to the thick position B, as described in Fig. 3A-D. Figure 3E shows the correlation factor $\kappa_{\mathrm{corr}}$ plotted as a function of the photon count rate. The obtained $\kappa_{\mathrm{corr}}$ values are larger than 1 unlike coherent CL of the previous section, showing super Poisson statistics of photons even within the excitation event by a single free electron. As predicted by Eq. 5, $\kappa_{\mathrm{corr}}$ is inversely proportional to the photon count rate, thus also to $\langle n^{\mathrm{med}} \rangle$. We excluded the thinnest two points with the lowest count rates for the fitting in Fig. 3E, which we discuss later. The asymptotic $\kappa_{\mathrm{corr}}$ value for a large count rate in Fig. 3E, corresponding to $\kappa_{\mathrm{corr}}^{\mathrm{med}}$ in Eq. 5, is $1.01 \pm 0.03$.

Thus, we can reasonably conclude that the statistics of the mediator particle generated within a single electron excitation event, i.e. $\kappa_{\mathrm{corr}}^{\mathrm{med}}$, follow the Poisson distribution, as has been previously reported for plasmons [37–40]. The Poisson statistics shown in these previous reports also imply that the mediator particles (c.f. plasmons) are the first particles excited by free electrons whereas an existence of additional particles exciting plasmons would lead to super-Poisson statistics of plasmons, which will be discussed in general in the next section.

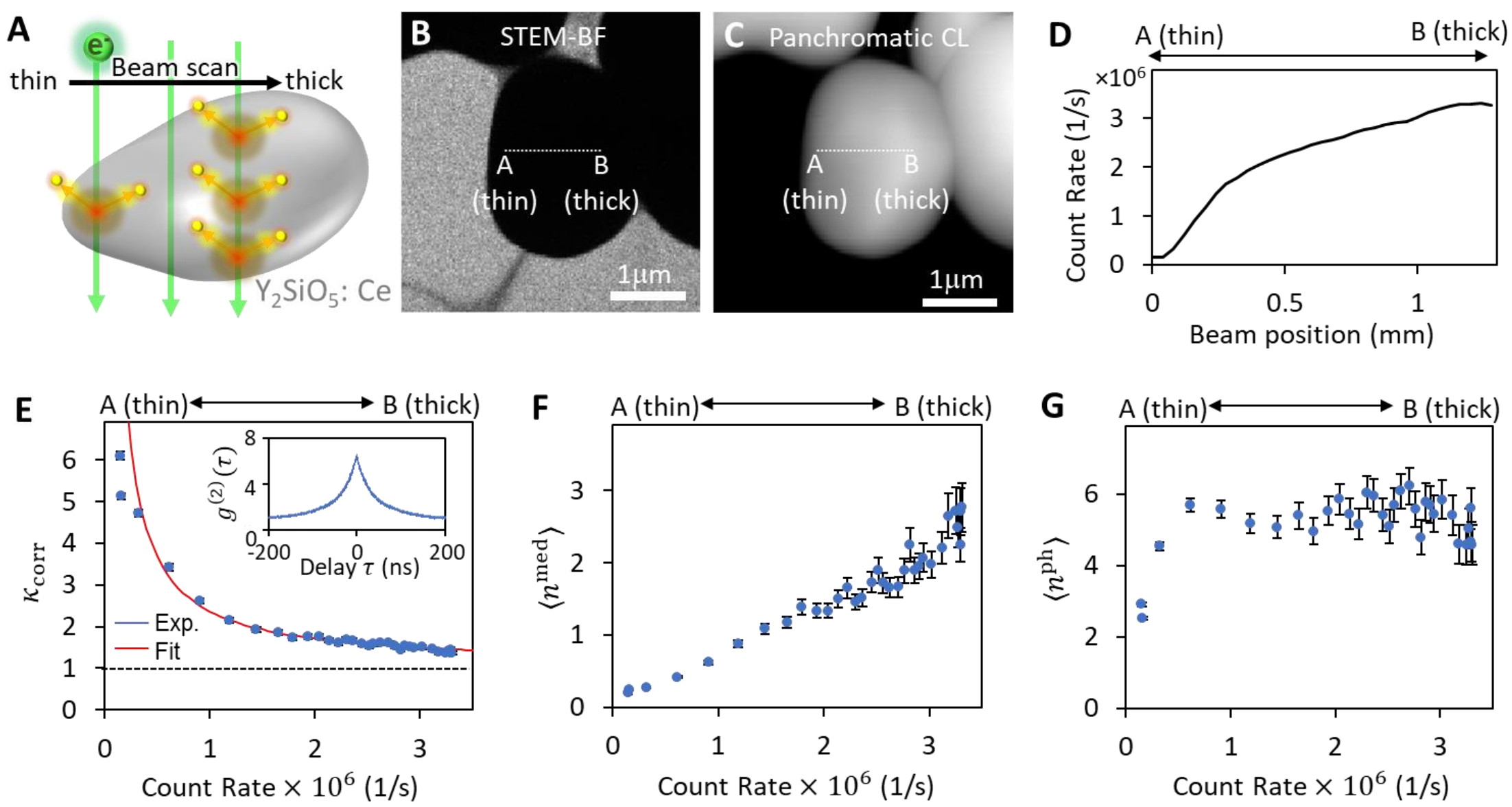

**Fig. 3. Statistical analysis results of incoherent cathodoluminescence.** (**A**) Schematic illustration of incoherent CL emission from YSO sample. (**B**) STEM bright field (BF) image and (**C**) panchromatic CL image of the measured sample. White dashed lines indicate the e-beam line scan trace from the thin position A to the thick position B. (**D**) A-B line profile of the photon count rate. (**E**) Correlation factor $\kappa_{\mathrm{corr}}$, (**F**) mean mediator particle number per electron $\langle n^{\mathrm{med}} \rangle$, and (**G**) mean photon number per mediator particle $\langle n^{\mathrm{ph}} \rangle_{1\mathrm{med}}$ plotted as functions of the count rate. The solid red line in (**E**) indicates the fitting line excluding the two left-most points. All the measurements were performed at an accelerating voltage of 80 kV and a beam current of 1.9 pA.

To estimate the average number of the mediator particle per electron $\langle n^{\mathrm{med}}\rangle$ from the experimentally obtained value of $\kappa_{\mathrm{corr}}$ using Eq. 5, we can sensibly assume that the generation of photons from a single mediator particle follows the Poisson statistics, i.e., $\kappa_{\mathrm{corr}}^{\mathrm{ph}} = 1$. Figure 3F shows the dependency of $\langle n^{\mathrm{med}}\rangle$ on the count rate, showing a linear trend. This proportionality originates from the process that a mediator particle produces $\langle n^{\mathrm{ph}}\rangle_{\mathrm{1med}}$ photons in average when a homogeneous sample can be assumed. One can also calculate the average number of photons excited by a single mediator particle $\langle n^{\mathrm{ph}}\rangle_{\mathrm{1med}}$ using the count rate being equal to $\eta \frac{I_{\mathrm{e}}}{e} \langle n^{\mathrm{med}}\rangle \langle n^{\mathrm{ph}}\rangle_{\mathrm{1med}}$, where $\eta$ ~2.2% is the detection efficiency of the system. Harnessing $\langle n^{\mathrm{med}}\rangle$ values from the correlation measurement, $\langle n^{\mathrm{ph}}\rangle_{\mathrm{1med}}$ is plotted along the beam scan in Fig. 3G, showing constant values independent of the count rate or thickness, except for the two thinnest (lowest count rate) positions. One candidate of possible mediator particles is bulk plasmons induced by incident free electrons [41]. Since the bulk plasmon has sufficiently higher energy than produced photons, a single plasmon can produce multiple photons while the mean number of photons per plasmon $\langle n^{\mathrm{ph}}\rangle_{\mathrm{1med}}$ should stay constant in the thick bulk regions without the surface effect. In contrast, in the two thinnest positions, or the lowest count rate positions, $\langle n^{\mathrm{ph}}\rangle_{\mathrm{1med}}$ is smaller than the other thicker areas, which can be explained by non-radiative relaxation (particle annihilation without producing photons) at the surface [42]. We note that additional effects, such as saturation of the photon source, interaction between mediator particles, or diffusion, could potentially alter the final photon statistics.

As shown here, the value of $\kappa_{\mathrm{corr}}$, which corresponds to the intensity correlation function $g^{(2)}(0)$ excluding the temporal modulation of a series of excitation events, exceeds one in incoherent CL. The mismatch between the obtained $\kappa_{\mathrm{corr}} > 1$ for CL and typical $g_{\mathrm{PL}}^{(2)}(0) = 1$ for PL[27] means that incoherent CL and PL show fundamentally different photon statistics. This indicates that photon statistics, which is the property of the stationary optical field, reflect nonequilibrium relaxation processes from the initial excitation to the final photon emission.

**Generalized photon statistics with multiple excitation steps**

On the basis of the above-mentioned description as well as the experimental verification, we here generalize the CL photon statistics using the correlation factor, even including the modulation by incident free electrons (Eq. 3). Equation 5 can explain that the excitation modulation by a mediator particle enhances the final correlation factor. This enhancement mechanism is actually similar to the excitation modulation by the free electrons as described previously [25]. If mediator excitation processes are multiple (cascade) and independent of each other ($i$-th step mediator only excites $i$+1-th step mediator), we can generalize the correlation factor as (see Supplementary Information for details):

$$g_{\mathrm{CL}}^{(2)}(0) = 1 + \frac{e}{2\tau_w I_{\mathrm{e}}}\left(\kappa_{\mathrm{corr}}^{(1)} + \frac{1}{\langle n^{(1)}\rangle}\left(\cdots\left(\kappa_{\mathrm{corr}}^{(\mathrm{m}-1)} + \frac{1}{\langle n^{(\mathrm{m}-1)}\rangle}\kappa_{\mathrm{corr}}^{\mathrm{m}}\right)\right)\right), \qquad (6)$$

where $\kappa_{\mathrm{corr}}^{(\mathrm{m})} = \frac{\left\langle \left(n^{(\mathrm{m})}\right)^2\right\rangle - \langle n^{(\mathrm{m})}\rangle}{\langle n^{(\mathrm{m})}\rangle^2}$ is the correlation factor of the $m$-th step particles excited by a single $(m-1)$-th step particle (Fig. 4). Considering the 0-th step particle as a fast electron and the final step particle as a photon, this model represents the general CL process. The first term of one corresponds to the Poissonian distribution of the electron, and the coefficient of the second term to the modulation by free electrons. The nested terms within the first parenthesis correspond to the previous $\kappa_{\mathrm{corr}}$ in Eq. 4 and Eq. 5, which is the pure photon statistics excluding the modulation by randomly incident free electrons. Then, for coherent CL, the first step particle ($m$ = 1) is a photon, while for incoherent CL with one additional mediating step the first step particle ($m$ = 1) is a mediator particle (e.g., plasmons) and the second step particle ($m$ = 2) is a photon. This expression universally describes the CL process both for coherent and incoherent processes and distinguishes the presence and absence of intermediate excitation processes. We note that the nested expression under the first parenthesis in Eq. 6 is a general formulation of particle statistics with multi-step (cascade) excitation processes.

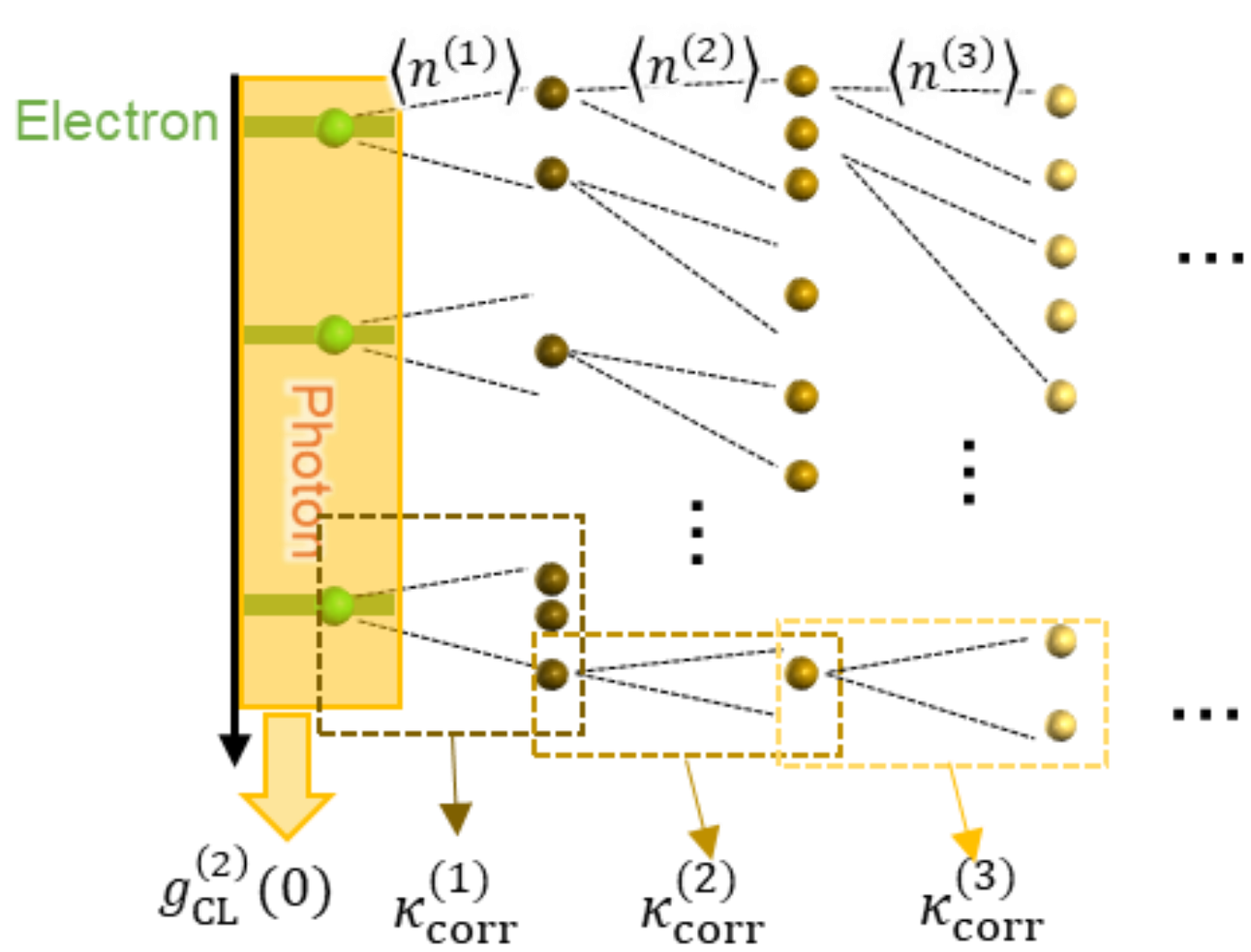

**Fig. 4. Schematic diagram of a cascade photon process with multiple excitation steps.** At $i$-th excitation step, average $\langle n^{(i)} \rangle$ particles are produced by a single ($i$-1)-th particle with the statistics represented by the correlation factor $\kappa_{\mathrm{corr}}^{(i)}$.

## Conclusion

We have demonstrated an approach to obtain the intrinsic photon statistics of CL photons within the individual electron excitation events by excluding the influence of excitation timing modulations in the e-beam. We experimentally evaluated the correlation factor, which describes pure photon statistics within a single electron excitation event, by varying several measurement parameters for both coherent and incoherent CL. The obtained results confirmed the theoretical description of photon statistics in coherent CL and clarified how mediator particles contribute to the photon statistics in incoherent CL, explaining the origin of the super-Poisson distribution. Thus, the proposed approach gives a comprehensive interpretation of the excitation mechanism of CL, which can also be potentially applied to PL with multi-step excitation processes. This study shows that the CL method not only offers higher spatial resolution than purely optical measurement but also provides insights into the intervening excitation processes prior to the light emission through photon statistics.

### References

1. Polman, A., Kociak, M. & García de Abajo, F. J. Electron-beam spectroscopy for nanophotonics. *Nat Mater* **18**, 1158–1171 (2019).
2. García De Abajo, F. J. & Di Giulio, V. Optical Excitations with Electron Beams: Challenges and Opportunities. *ACS Photonics* **8**, 945–974 (2021).
3. Dang, Z., Chen, Y. & Fang, Z. Cathodoluminescence nanoscopy: state of the art and beyond. *ACS Nano* **17**, 24431–24448 (2023).
4. García De Abajo, F. J. Optical excitations in electron microscopy. *Rev Mod Phys* **82**, 209–275 (2010).
5. Dmitruk, N. L., Litovchenko, V. G. & Talat, G. H. The effect of the surface space charge region on the cathodoluminescence of semiconductors. *Surf Sci* **72**, 321–341 (1978).
6. Boulou, M. & Bois, D. Cathodoluminescence measurements of the minority-carrier lifetime in semiconductors. *J Appl Phys* **48**, 4713–4721 (1977).
7. Yacobi, B. G. & Holt, D. B. Cathodoluminescence scanning electron microscopy of semiconductors. *J Appl Phys* **59**, (1986).
8. Guthrey, H. & Moseley, J. A Review and Perspective on Cathodoluminescence Analysis of Halide Perovskites. *Adv Energy Mater* **10**, (2020).
9. Ozawa, L. & Itoh, M. Cathode Ray Tube Phosphors. *Chem Rev* **103**, 3835–3855 (2003).
10. Kociak, M. & Stéphan, O. Mapping plasmons at the nanometer scale in an electron microscope. *Chem. Soc. Rev.* **43**, 3865–3883 (2014).
11. Van Wijngaarden, J. T. *et al.* Direct imaging of propagation and damping of near-resonance surface plasmon polaritons using cathodoluminescence spectroscopy. *Appl Phys Lett* **88**, (2006).

12. Bashevoy, M. V. *et al.* Generation of traveling surface plasmon waves by free-electron impact. *Nano Lett* **6**, 1113–1115 (2006).
13. Feist, A. *et al.* Cavity-mediated electron-photon pairs. *Science (1979)* **377**, 777–780 (2022).
14. Yanagimoto, S. *et al.* Time-correlated electron and photon counting microscopy. *Commun Phys* **6**, (2023).
15. Huang, G., Engelsen, N. J., Kfir, O., Ropers, C. & Kippenberg, T. J. Electron-Photon Quantum State Heralding Using Photonic Integrated Circuits. *PRX Quantum* **4**, (2023).
16. Karnieli, A. & Fan, S. Jaynes-Cummings interaction between low-energy free electrons and cavity photons. *Science (1979)* **9**, (2023).
17. Dang, Z., Chen, Y. & Fang, Z. Cathodoluminescence nanoscopy: state of the art and beyond. *ACS Nano* **17**, 24431–24448 (2023).
18. Roques-Carmes, C. *et al.* Free-electron-light interactions in nanophotonics. *Appl Phys Rev* **10**, (2023).
19. Coenen, T. & Haegel, N. M. Cathodoluminescence for the 21st century: Learning more from light. *Appl Phys Rev* **4**, (2017).
20. Scheucher, M., Schachinger, T., Spielauer, T., Stöger-Pollach, M. & Haslinger, P. Discrimination of coherent and incoherent cathodoluminescence using temporal photon correlations. *Ultramicroscopy* **241**, 113594 (2022).
21. Sola-Garcia, M. *et al.* Photon Statistics of Incoherent Cathodoluminescence with Continuous and Pulsed Electron Beams. *ACS Photonics* **8**, 916–925 (2021).
22. Kubota, T. *et al.* Cathodoluminescence spectral and lifetime mapping of $Cs_4PbBr_6$ : fast lifetime and its scintillator application. *Applied Physics Express* **17**, 015005 (2024).
23. Fiedler, S. *et al.* Photon superbunching in cathodoluminescence of excitons in WS2 monolayer. *2d Mater* **10**, (2023).
24. Iyer, V. *et al.* Photon bunching in cathodoluminescence induced by indirect electron excitation. *Nanoscale* **15**, 9738–9744 (2023).
25. Yuge, T., Yamamoto, N., Sannomiya, T. & Akiba, K. Superbunching in cathodoluminescence : A master equation approach. *Phys Rev B* **107**, 165303 (2023).
26. Yanagimoto, S., Yamamoto, N., Sannomiya, T. & Akiba, K. Purcell effect of nitrogen-vacancy centers in nanodiamond coupled to propagating and localized surface plasmons revealed by photon-correlation cathodoluminescence. *Phys Rev B* **103**, 205418 (2021).
27. Meuret, S. *et al.* Photon bunching in cathodoluminescence. *Phys Rev Lett* **114**, (2015).
28. Tizei, L. H. G. & Kociak, M. Spatially resolved quantum nano-optics of single photons using an electron microscope. *Phys Rev Lett* **110**, 1–5 (2013).
29. García de Abajo, F. J. Multiple excitation of confined graphene plasmons by single free electrons. *ACS Nano* **7**, 11409–11419 (2013).
30. Yamamoto, N., Araya, K. & García de Abajo, F. J. Photon emission from silver particles induced by a high-energy electron beam. *Phys Rev B Condens Matter Mater Phys* **64**, 2054191–2054199 (2001).
31. Gersten, J. & Nitzan, A. Spectroscopic properties of molecules interacting with small dielectric particles. *J Chem Phys* **75**, 1139–1152 (1981).
32. Brenny, B. J. M., Polman, A. & García de Abajo, F. J. Femtosecond plasmon and photon wave packets excited by a high-energy electron on a metal or dielectric surface. *Phys Rev B* **94**, 1–11 (2016).
33. Yamamoto, N., García de Abajo & Myroshnychenko, V. Interference of surface plasmons and Smith-Purcell emission probed by angle-resolved cathodoluminescence spectroscopy. *Phys Rev B* **91**, 125144 (2015).
34. Smiths, S. J., Purcell, E. M. & Kursunoglu, B. Visible Light from Localized Surface Charges Moving across a Grating Derivation and Renormalization of the Tamm-Dancoff Equations't. *Physical Review* **92**, 1069 (1953).
35. Yamamoto, N., Sugiyama, H. & Toda, A. Cherenkov and transition radiation from thin plate crystals detected in the transmission electron microscope. *Proceedings of the Royal Society A: Mathematical, Physical and Engineering Sciences* **452**, 2279–2301 (1996).
36. Meuret, S. *et al.* Photon bunching reveals single-electron cathodoluminescence excitation efficiency in InGaN quantum wells. *Phys Rev B* **96**, 1–8 (2017).
37. Pines, D. Collective energy losses in solids. *Rev Mod Phys* **28**, 184 (1956).
38. Shockley, W. Problems related to p-n junctions in silicon. *Solid State Electronics* **2**, (1961).

39. Egerton, R. F. *Electron Energy-Loss Spectroscopy in the Electron Microscope*. *Electron Energy-Loss Spectroscopy in the Electron Microscope* (Springer US, 2011). doi:10.1007/978-1-4419-9583-4.
40. Rothwarf, A. Plasmon theory of electron-hole pair production: Efficiency of cathode ray phosphors. *J Appl Phys* **44**, 752–756 (1973).
41. Varkentina, N. *et al.* Cathodoluminescence excitation spectroscopy: Nanoscale imaging of excitation pathways. *Sci Adv* **8**, 1–8 (2022).
42. Van Dijken, A., Meulenkamp, E. A., Vanmaekelbergh, D. & Meijerink, A. The Kinetics of the Radiative and Nonradiative Processes in Nanocrystalline ZnO Particles upon Photoexcitation. *Journal of Physical Chemistry B* **104**, 1715–1723 (2000).

**Acknowledgments:** The authors are grateful toProf. Kunio Takayanagi for fruitful discussions and advice. We gratefully acknowledge Dr. Takeshi Ohshima for his kind support of the research. This work is financially supported by JSPS KAKENHI JP21K18195 (T.S), JP22H01963 (K.A.), JP22H05032 (K.A.), JP24H00400(T.S.), JST FOREST JPMJFR213J (T.S.) and JSPS PD2(20J14821) (S.Y.).

**Author contributions:**

Conceptualization: K.A., T.S.

Methodology: S.Y., N.Y., T.Y., K.A., T.S.

Investigation: S.Y., N.Y., T.Y., K.A., T.S.

Visualization: S.Y., K.A., T.S.

Funding acquisition: K.A., T.S.

Project administration: K.A., T.S.

Supervision: K.A., T.S.

Writing – original draft: S.Y.

Writing – review & editing: S.Y., N.Y., T.Y., K.A., T.S.

**Competing interests:** Authors declare that they have no competing interests.

## Methods

### Instrumentation

A scanning transmission electron microscope JEM 2000FX (JEOL., Japan) equipped with a parabolic mirror above the sample to collected the light emission was used for the measurement. (Fig. 1A) The light beam from the sample is split into two paths in free space using a 50:50 beam splitter and then coupled into two optical fibers, which are connected to single-photon counting modules, $D_A$ and $D_B$ (PDM, MPD, Italy). The histogram $H(\tau)$ (see Fig. 1B) of coincidence events between $D_A$ and $D_B$ is acquired by using the electric correlator (TimeHarp260 PICO, Pico Quant, Germany). The photon detection efficiency of our system is estimated to be ~2.20% (see also Supplementary Information). We performed all measurements under conditions where the light intensity was sufficient compared to the noise signal (in the order of $10^2$ counts per second) and set the integration time depending on the signal strength. In the case of transition radiation shown in Fig. 2, which exhibits relatively low intensities compared to the other coherent CL, the count rates for each detector at 120 kV, 160 kV, and 200 kV were $1.2 \times 10^4$, $1.9 \times 10^4$, and $3.0 \times 10^4$ counts per second, respectively, with the integration times of 7500 s, 4000 s, and 3200 s.

**Photon statistics and second order correlation in CL**

We here derive the relation of $\kappa_{\mathrm{corr}}$ and $g^{(2)}$ and their expressions used in the main text. The Hanbury Brown-Twiss (HBT) method enables measuring the second order correlation function of intensity $I$, $g^{(2)}(\tau) = \frac{\langle I(t)I(t+\tau)\rangle}{\langle I(t)\rangle\langle I(t+\tau)\rangle} = \frac{\langle n_{\mathrm{A}}(t)n_{\mathrm{B}}(t+\tau)\rangle}{\langle n_{\mathrm{A}}(t)\rangle\langle n_{\mathrm{B}}(t+\tau)\rangle}$, where $n_{\mathrm{A(B)}}$ is the number of photons introduced into the detector $D_{A(B)}$ at time $t$, $\tau$ is time delay and $\langle\cdots\rangle$ means ensemble average.(Fig. 1A) When the integration time of the statistical data is sufficiently long, the time averaging can be treated as ensemble averaging according to the ergodic theorem, leading to $g^{(2)}(\tau) = \frac{H(\tau)}{H(\tau\to\infty)}$. The coincidence counting event with time delay $\tau$ in the CL-HBT measurement can be classified into two cases: (i) coincidence counting of photons excited by different electrons and (ii) coincidence counting of photons excited by the same electron.

Firstly, we consider case (i). The delay of photon detection events excited by different electrons is affected by the correlation of electron arrival times. When the delay is sufficiently longer than the electron coherence time, the electron arrival events are uncorrelated, and therefore, these events contribute to the flat background shown in the blue shaded area of the histogram in Fig. 1B. The background height $h_{\mathrm{uncorr}}$ is same also for the zero delay $\tau = 0$ because the coherence time ($\tau_c$~fs) of the thermal electron gun used in this study is significantly shorter than the time resolution $\sim$500ps of our correlation measurement setup (see the section of Instrument Response Function in the Supplementary Information).

Now we define $p_{\mathrm{n}}$ as the probability of $n$ photons being emitted by one electron incident event. When the emitted $n$ photons are split by the beam splitter, $k$ and $n-k$ photons are delivered into detectors $D_A$ and $D_B$, respectively. Then the average photon number in these detectors $\langle n_{\mathrm{A}}\rangle$ and $\langle n_{\mathrm{B}}\rangle$ are derived as:

$$\langle n_{\mathrm{A}}\rangle = \eta_{\mathrm{A}}\frac{I_{\mathrm{e}}}{e}\left(\sum_{n=0}^{\infty} p_{\mathrm{n}} \sum_{k=0}^{n} \frac{{}_{n}C_{k}}{2^{n}} k\right) = \frac{1}{2}\eta_{\mathrm{A}}\frac{I_{\mathrm{e}}}{e}\sum_{n=0}^{\infty} p_{\mathrm{n}}\, n \ ,$$

$$\langle n_{\mathrm{B}}\rangle = \eta_{\mathrm{B}}\frac{I_{\mathrm{e}}}{e}\left(\sum_{n=0}^{\infty} p_{\mathrm{n}} \sum_{k=0}^{n} \frac{{}_{n}C_{k}}{2^{n}} (n-k)\right) = \frac{1}{2}\eta_{\mathrm{B}}\frac{I_{\mathrm{e}}}{e}\sum_{n=0}^{\infty} p_{\mathrm{n}}\, n \ , \tag{7}$$

where $I_{\mathrm{e}}$ is electron beam current, $e$ is elementary charge, and $\eta_{\mathrm{A(B)}}$ is overall detection efficiency of the system. Also, we utilize the relation $\sum_{k=0}^{n} \frac{{}_{n}C_{k}}{2^{n}} k = \sum_{k=0}^{n} \frac{{}_{n}C_{k}}{2^{n}} (n-k) = \frac{n}{2}$. Using the total coincidence count $\langle n_{\mathrm{A}}\rangle T \cdot \langle n_{\mathrm{B}}\rangle T$ within the integration time $T$ and the number of bins $\frac{T}{t_{\mathrm{bin}}}$, the uncorrelated count $h_{\mathrm{uncorr}}$ in binning time $t_{\mathrm{bin}}$ can be expressed as

$$h_{\mathrm{uncorr}} = \frac{\langle n_{\mathrm{A}}\rangle T \langle n_{\mathrm{B}}\rangle T}{\frac{T}{t_{\mathrm{bin}}}} = \frac{1}{4}\eta_{\mathrm{A}}\eta_{\mathrm{B}}\left(\frac{I_{\mathrm{e}}}{e}\right)^{2} T t_{\mathrm{bin}} \left(\sum_{n=0}^{\infty} p_{\mathrm{n}}\, n\right)^{2} . \tag{8}$$

This $h_{\mathrm{uncorr}}$ corresponds to the flat background count, as shown as blue-shaded region with $H(\tau) \leq h_{\mathrm{uncorr}}$ in Fig. 1B.

Secondly, we consider the case (ii) (shaded area in the histogram in Fig. 1B, the coincidence of the photons generated by the same electron. The total number of coincident count of photons excited by the same electron is described as

$$N_{\mathrm{corr}} = \eta_{\mathrm{A}}\eta_{\mathrm{B}} T \frac{I_{\mathrm{e}}}{e} \sum_{n=0}^{\infty} p_{\mathrm{n}} \sum_{k=0}^{n} \frac{{}_{n}C_{k}}{2^{n}} k(n-k) = \frac{1}{4}\eta_{\mathrm{A}}\eta_{\mathrm{B}} T \frac{I_{\mathrm{e}}}{e} \sum_{n=0}^{\infty} p_{\mathrm{n}} n(n-1) . \tag{9}$$

Photons excited by one electron are observed within time $\tau_{\mathrm{w}}$, which consists of the lifetime of photon source and the instrument response time. The total coincident count $N_{\mathrm{corr}}$ is spread within the typical correlation time width of $\tau_{\mathrm{w}}$. In the histogram of Fig. 1B, the integrated count higher than the flat background (the red-shaded area ($H(\tau) > h_{\mathrm{uncorr}}$) around $\tau = 0$ with the width of approximately $\tau_{\mathrm{w}}$) corresponds to the coincident count $N_{\mathrm{corr}}$. Here, we introduce a shape factor $\beta (> 0)$ to describe $h_{\mathrm{corr}}$ , which represents the height of the red-shaded area in Fig. 1B, as follows:

$$h_{\mathrm{corr}} = \frac{N_{\mathrm{corr}} t_{\mathrm{bin}}}{2\beta\tau_{\mathrm{w}}} . \tag{10}$$

Using Eq. 8-10, the peak value at $\tau = 0$ of the correlation function of CL becomes

$$g_{\mathrm{CL}}^{(2)}(0) = \frac{h_{\mathrm{uncorr}} + h_{\mathrm{corr}}}{h_{\mathrm{uncorr}}} = 1 + \frac{e}{2\beta I_{\mathrm{e}}\tau_{\mathrm{w}}} \kappa_{\mathrm{corr}} , \tag{11}$$

where the correlation factor $\kappa_{\text{corr}} = \frac{\sum_{n=0}^{\infty} p_{\text{n}} n(n-1)}{\left(\sum_{n=0}^{\infty} p_{\text{n}} n\right)^2} = \frac{\langle n(n-1)\rangle_{1\text{e}}}{\langle n\rangle_{1\text{e}}{}^2} = \frac{\langle n^2\rangle_{1\text{e}} - \langle n\rangle_{1\text{e}}}{\langle n\rangle_{1\text{e}}{}^2}$ is introduced. The suffix "1e" represents averaging within a single electron excitation event. For the typical time decay of an exponential function, $\beta = 1$ can be reasonably assumed[21]. The coefficient of $\kappa_{\text{corr}}$ (the factor of $\frac{e}{2\beta I_{\text{e}} \tau_{\text{w}}}$) in Eq. 5 can be understood as the temporal modulation of a series of excitation events, which consequently modulates the photon statistics. $\kappa_{\text{corr}}$ provides the statistical fluctuation in the number of emitted photons during individual excitation events. In other words, $\kappa_{\text{corr}}$ corresponds to the intensity correlation of photons, or pure photon statistics, within the excitation event by a single electron. When the photon statistics follows the Poisson distribution, $\kappa_{\text{corr}} = 1$ holds. According to the previously reported expression[26], we can write the time dependent expression with exponential decay ($\beta = 1$) as,

$$g_{\text{CL}}^{(2)}(\tau) = 1 + \frac{e}{2I_{\text{e}}\tau_{\text{w}}} \kappa_{\text{corr}} \exp\left(-\frac{|\tau|}{\tau_w}\right). \tag{12}$$

Supplementary Information for

# Unveiling the Nature of Cathodoluminescence from Photon Statistics

Sotatsu Yanagimoto, Naoki Yamamoto, Tatsuro Yuge, Takumi Sannomiya, Keiichirou Akiba

Corresponding authors:
Takumi Sannomiya, sannomiya.t.aa@m.titech.ac.jp
Keiichirou Akiba, akiba.keiichiro@qst.go.jp

## Photon Statistics in Coherent CL

In the coherent CL process, a free electron directly excites photons [1]. This process can be described using the scattering operator $\hat{S} = \exp\left(g\hat{b}\hat{a}^\dagger - g^*\hat{b}^\dagger\hat{a}\right)$ [2], where $g$ is a coupling strength, $\hat{b}^\dagger$ and $\hat{b}$ are electron energy ladder operators, and $\hat{a}^\dagger$ and $\hat{a}$ are photon creation and annihilation operators, respectively. Provided that $\hat{b}^\dagger$ and $\hat{b}$ are treated as c-numbers [3], the scattering operator $\hat{S}$ behaves as a displacement operator $\hat{D}$ [4]. Then, the coherent CL process generates a coherent state of light $|\alpha\rangle = \exp\left(-\frac{|\alpha|^2}{2}\right)\sum_n \frac{\alpha^n}{\sqrt{n!}}|n\rangle$, where $|n\rangle$ is the Fock state and $\alpha$ is a complex number. The photon statistics of this state follows the Poisson distribution, i.e., $\kappa_{\mathrm{corr}} = 1$. Even if $\hat{b}^\dagger$ and $\hat{b}$ cannot be treated as c-numbers (e.g. the photon and electron are entangled), the diagonal elements of the reduced density matrix of the photon system are unchanged, meaning that the photon statistics still follows the Poisson distribution in coherent CL. Thus, the correlation factor of coherent CL is expected to equal one ($\kappa_{\mathrm{corr}} = 1$) regardless of the mean photon number $\langle n\rangle$.

## Photon Statistics in Incoherent CL

We analytically describe the incoherent CL photon statistics within a single electron excitation event using generation probabilities of photocarriers. In the incoherent CL, photons are excited through mediator particles, such as bulk plasmons or secondary electrons, thus involving relaxation processes [5]. As the simplest model, we consider two steps in the excitation process: (1) fast electrons generate mediator particles and (2) the mediator particles generate photons. Suppose that a single electron excites $n^{\mathrm{med}}$ mediator particles and the $m$-th mediating particle excites $n_m^{\mathrm{ph}}$ photons. We define the excitation probabilities of the mediator particles and photons as $P^{\mathrm{med}}(n^{\mathrm{med}})$ and $P^{\mathrm{ph}}(n^{\mathrm{ph}})$, respectively. Assuming that the excitations by individual mediator particles are independent, we can calculate the mean (ensemble average) $\langle n^{\mathrm{ph}}\rangle_{n^{\mathrm{med}}}$ and mean square of photons $\left\langle \left(n^{\mathrm{ph}}\right)^2\right\rangle_{n^{\mathrm{med}}}$ by $n^{\mathrm{med}}$ mediator particles as:

$$\langle n^{\mathrm{ph}}\rangle_{n^{\mathrm{med}}} = \sum_{m=1}^{n^{\mathrm{med}}} \sum_{n_m^{\mathrm{ph}}=0}^{\infty} n_m^{\mathrm{ph}} \cdot P^{\mathrm{ph}}(n_m^{\mathrm{ph}}) = n^{\mathrm{med}} \cdot \langle n^{\mathrm{ph}}\rangle_{1\mathrm{med}}$$

$$\begin{aligned}\left\langle \left(n^{\mathrm{ph}}\right)^2\right\rangle_{n^{\mathrm{med}}} &= \sum_{m\neq m'}^{n^{\mathrm{med}}} \sum_{n_m^{\mathrm{ph}}=0}^{\infty} n_m^{\mathrm{ph}} \cdot P^{\mathrm{ph}}\left(n_m^{\mathrm{ph}}\right) \sum_{n_{m'}^{\mathrm{ph}}=0}^{\infty} n_{m'}^{\mathrm{ph}} \cdot P^{\mathrm{ph}}\left(n_{m'}^{\mathrm{ph}}\right) \\ &+ \sum_{m=0}^{n^{\mathrm{med}}} \sum_{n_m^{\mathrm{ph}}=0}^{\infty} \left(n_m^{\mathrm{ph}}\right)^2 \cdot P^{\mathrm{ph}}\left(n_m^{\mathrm{ph}}\right) \\ &= n^{\mathrm{med}}\left(n^{\mathrm{med}}-1\right)\langle n^{\mathrm{ph}}\rangle_{1\mathrm{med}}^2 + n^{\mathrm{med}}\left\langle \left(n^{\mathrm{ph}}\right)^2\right\rangle_{1\mathrm{med}},\end{aligned} \tag{S1}$$

where $\langle n^{\mathrm{ph}}\rangle_{1\mathrm{med}} = \sum_{n_m^{\mathrm{ph}}=0}^{\infty} n_m^{\mathrm{ph}} \cdot P^{\mathrm{ph}}(n_m^{\mathrm{ph}})$ is the mean number of photons generated by a single mediator particle and $\left\langle \left(n^{\mathrm{ph}}\right)^2\right\rangle_{1\mathrm{med}} = \sum_{n_m^{\mathrm{ph}}=0}^{\infty}\left(n_m^{\mathrm{ph}}\right)^2 \cdot P^{\mathrm{ph}}(n_m^{\mathrm{ph}})$. By averaging these values over $n^{\mathrm{med}}$, we obtain

$$\langle n^{\mathrm{ph}}\rangle_{1\mathrm{e}} = \sum_{n^{\mathrm{med}}=0}^{\infty} \langle n^{\mathrm{ph}}\rangle_{n^{\mathrm{med}}} \cdot P^{\mathrm{med}}\left(n^{\mathrm{med}}\right) = \langle n^{\mathrm{med}}\rangle_{1\mathrm{e}} \cdot \langle n^{\mathrm{ph}}\rangle_{1\mathrm{med}}$$

$$\begin{aligned}\left\langle \left(n^{\mathrm{ph}}\right)^2\right\rangle_{1\mathrm{e}} &= \sum_{n^{\mathrm{med}}=0}^{\infty} \left\langle \left(n^{\mathrm{ph}}\right)^2\right\rangle_{n^{\mathrm{med}}} \cdot P^{\mathrm{med}}\left(n^{\mathrm{med}}\right) \\ &= \left(\left\langle \left(n^{\mathrm{med}}\right)^2\right\rangle_{1\mathrm{e}} - \langle n^{\mathrm{med}}\rangle_{1\mathrm{e}}\right)\langle n^{\mathrm{ph}}\rangle_{1\mathrm{med}}^2 + \langle n^{\mathrm{med}}\rangle_{1\mathrm{e}} \cdot \left\langle \left(n^{\mathrm{ph}}\right)^2\right\rangle_{1\mathrm{med}},\end{aligned} \tag{S2}$$

where $\langle n^{\mathrm{med}}\rangle_{1\mathrm{e}} = \sum_{n^{\mathrm{med}}=0}^{\infty} n^{\mathrm{med}} \cdot P^{\mathrm{med}}(n^{\mathrm{med}})$ is the mean number of mediator particles generated by a single electron and $\left\langle \left(n^{\mathrm{med}}\right)^2\right\rangle_{1\mathrm{e}} - \langle n^{\mathrm{med}}\rangle_{1\mathrm{e}} = \sum_{n^{\mathrm{med}}=0}^{\infty}\left(\left(n^{\mathrm{med}}\right)^2 - n^{\mathrm{med}}\right) \cdot$ $P^{\mathrm{med}}(n^{\mathrm{med}})$. We note that $\langle n^{\mathrm{med}}\rangle_{1\mathrm{e}}$ is expressed as $\langle n^{\mathrm{med}}\rangle$ without suffix "1e" in the main text for the ease of reading. Considering $\kappa_{\mathrm{corr}} = \left(\left\langle (n^{\mathrm{ph}})^2 \right\rangle_{1\mathrm{e}} - \langle n^{\mathrm{ph}}\rangle_{1\mathrm{e}}\right) / \langle n^{\mathrm{ph}}\rangle_{1\mathrm{e}}{}^2$ and Eq. (S1) and (S2), the correlation factor for incoherent CL is described as:

$$\kappa_{\mathrm{corr}} = \kappa_{\mathrm{corr}}^{\mathrm{med}} + \frac{1}{\langle n^{\mathrm{med}}\rangle_{1\mathrm{e}}} \kappa_{\mathrm{corr}}^{\mathrm{ph}} \ , \tag{S3}$$

where $\kappa_{\mathrm{corr}}^{\mathrm{med}} = \frac{\left\langle (n^{\mathrm{med}})^2 \right\rangle_{1\mathrm{e}} - \langle n^{\mathrm{med}}\rangle_{1\mathrm{e}}}{\langle n^{\mathrm{med}}\rangle_{1\mathrm{e}}{}^2}$ and $\kappa_{\mathrm{corr}}^{\mathrm{ph}} = \frac{\left\langle (n^{\mathrm{ph}})^2 \right\rangle_{1\mathrm{med}} - \langle n^{\mathrm{ph}}\rangle_{1\mathrm{med}}}{\langle n^{\mathrm{ph}}\rangle_{1\mathrm{med}}{}^2}$. Assuming that each excitation probability follows the Poisson distribution, $\kappa_{\mathrm{corr}}^{\mathrm{med}} = \kappa_{\mathrm{corr}}^{\mathrm{ph}} = 1$, we can rewrite Eq. (S3) as:

$$\kappa_{\mathrm{corr}} = 1 + \frac{1}{\langle n^{\mathrm{med}}\rangle_{1\mathrm{e}}}. \tag{S4}$$

In incoherent CL, including successive multiple excitation processes, the correlation factor $\kappa_{\mathrm{corr}}$ exceeds one, being enhanced by the mean number (ensemble average per electron) of mediator particles $\langle n^{\mathrm{med}}\rangle_{1\mathrm{e}}$. Thus, $\kappa_{\mathrm{corr}}$ is expected to provide insights into the photon generation processes and enables us to distinguish the coherent and incoherent CL.

Since the expression of Eq. (S3) can be applied multiple times, i.e., $\kappa_{\mathrm{corr}}^{\mathrm{ph}}$ in Eq. (S3) contains another excitation process, we can extend this approach to the multi-step excitation process with nested terms:

$$\kappa_{\mathrm{corr}} = \kappa_{\mathrm{corr}}^{(1)} + \frac{1}{\langle n^{(1)}\rangle}\left(\kappa_{\mathrm{corr}}^{(2)} + \frac{1}{\langle n^{(2)}\rangle} \cdots \left(\kappa_{\mathrm{corr}}^{(\mathrm{m}-1)} + \frac{1}{\langle n^{(\mathrm{m}-1)}\rangle} \kappa_{\mathrm{corr}}^{\mathrm{m}}\right)\right). \tag{S5}$$

$\kappa_{\mathrm{corr}}^{(\mathrm{m})} = \frac{\left\langle (n^{(\mathrm{m})})^2 \right\rangle - \langle n^{(\mathrm{m})}\rangle}{\langle n^{(\mathrm{m})}\rangle^2}$ is the correlation factor of the $m$-th step particles excited by a single $(m-1)$-th step particle. This expression corresponds to the term of the parenthesis in Eq. 6 in the main text.

## Instrument Response Function

The shape of the correlation curve reflects the physical properties of the sample and the instrument response function (IRF). By performing the HBT measurement of ultrafast pulsed laser with a pulse width of 50 fs, we can evaluate the IRF. Figure S1 shows results with the ultrafast pulse laser at two wavelengths, 400 and 800 nm. By fitting with a double Gaussian function,

$$g^{(2)}(\tau) = a_1 \exp\left(-\left(\frac{\tau}{\tau_1}\right)^2\right) + a_2 \exp\left(-\left(\frac{\tau}{\tau_2}\right)^2\right) + 1, \quad (S6)$$

we obtained the decay time $\tau_1 = 0.056$ ns and $\tau_2 = 0.460$ ns for the wavelength of 400 nm, and $\tau_1 = 0.058$ ns, $\tau_2 = 0.453$ ns for the wavelength of 800 nm, as shown in Fig. S1. For instance, the Cherenkov radiation (coherent CL) from mica at the acceleration voltage of 200 kV shows correlation time of $\tau_w = 0.495$ ns, which is governed by the decay $\tau_2$. The small increase of the correlation time from the IRF is related to the pulse shape deterioration in the electric signal due to the longer electrical cables used in the CL-HBT measurement compared to the IRF measurement.

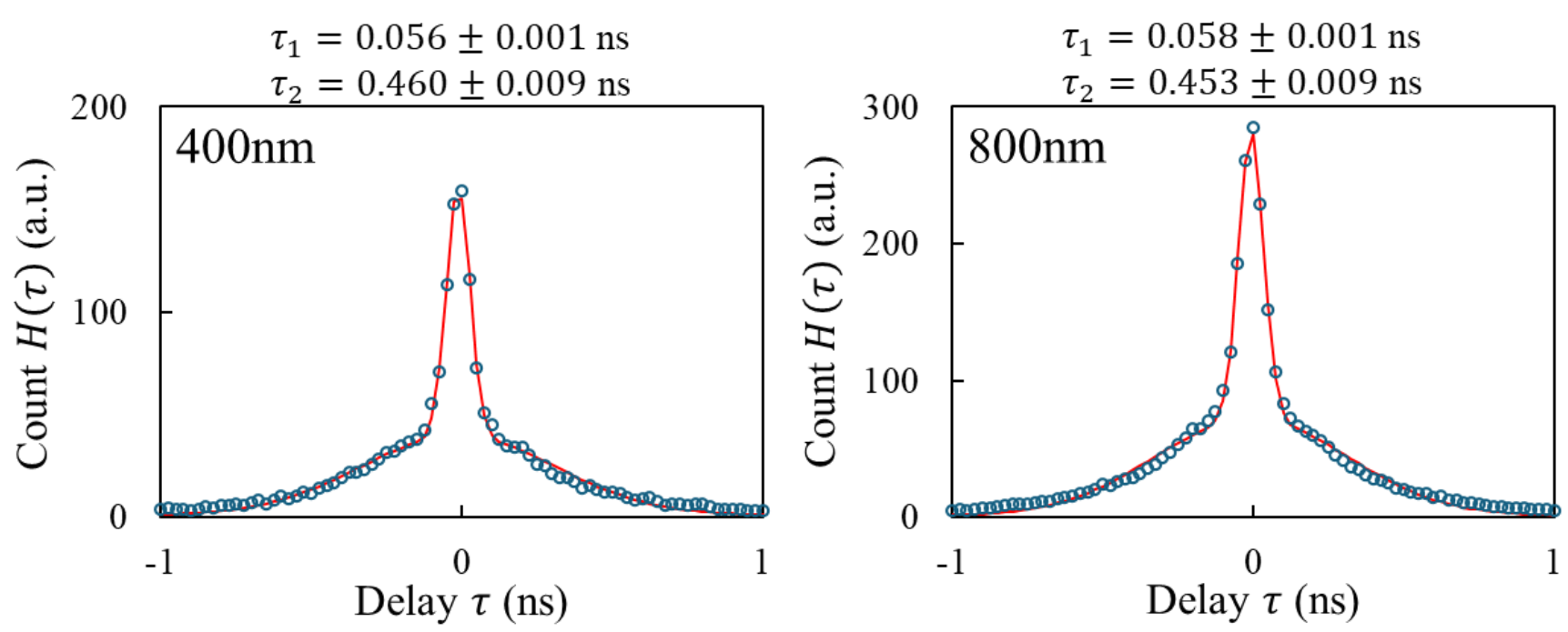

**Fig. S1. Instrument response function.** Correlation curves measured with femto-second pulsed laser at two wavelengths, namely (a) 400 nm and (b) 800 nm. The obtained data are fitted with a double Gaussian function expressed by Eq. (S6).

Detection Efficiency

The detection efficiency of the CL measurement system can be estimated from the measurement of transition radiation (TR), which can be theoretically calculated [5]. Considering the detection solid angle $\Omega$ by the parabolic mirror as a half of the upper hemisphere ($\Omega = \pi$), the excitation probability of TR in the wavelength range of 300-900 nm by a single electron is shown in Fig. S2 for different acceleration voltages. The theoretical excitation probability of TR at 200 kV is $1.06 \times 10^{-3}$ [photons/electron], as shown in Fig. S2. Using the experimentally obtained photon count rate of $6 \times 10^{-4}$ [photons/s] at 200 kV and the electron beam current of 400 pA, a detection efficiency of 2.2% is obtained [6].

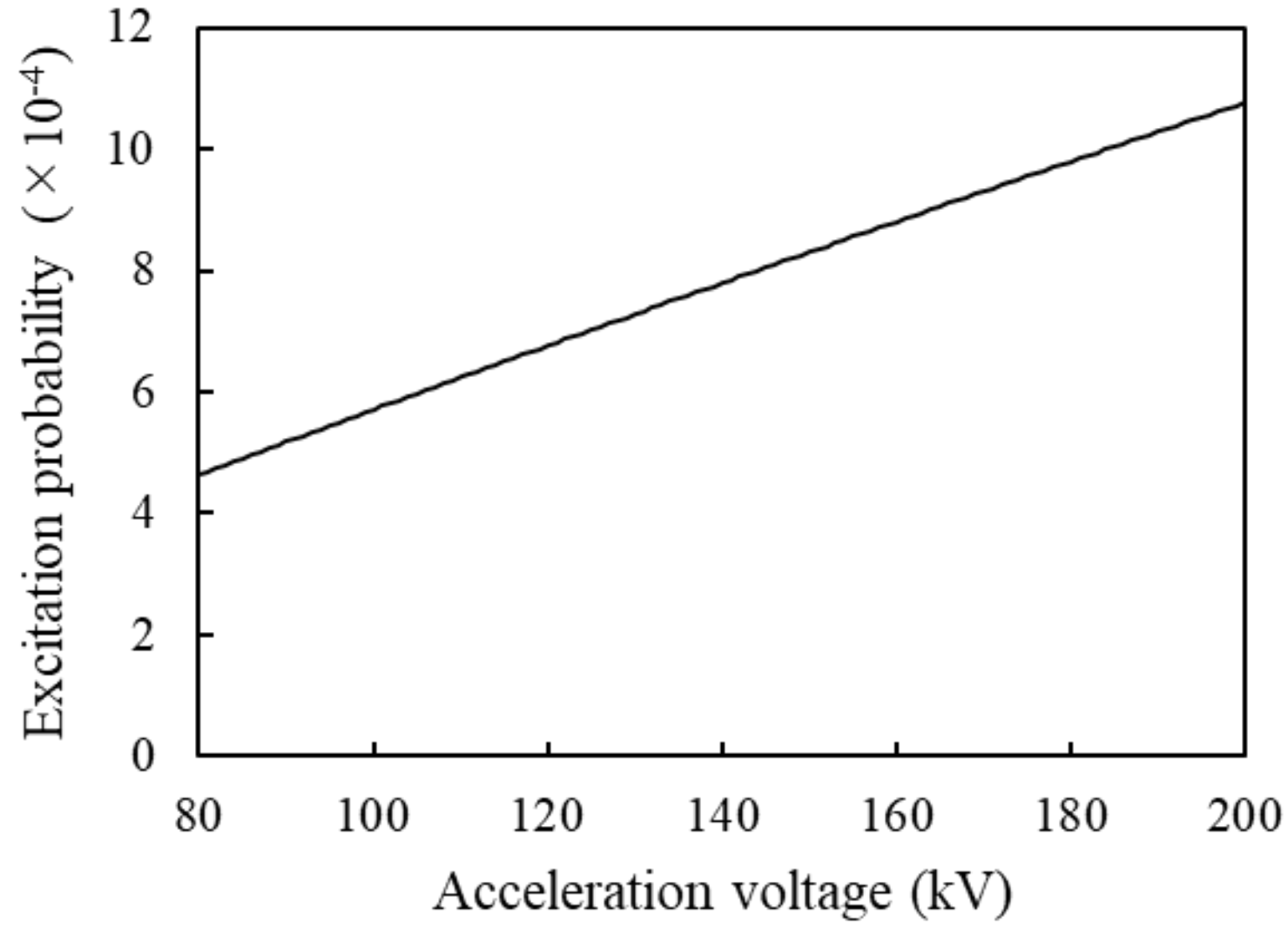

**Fig. S2. Calculated excitation probability of TR.** The dielectric constant of silver is given by Palik [7]. The probability values are integrated over the wavelength range of 300 to 900 nm.

Coherent CL from Flat Silver Surface Connected to Periodic Structure

We evaluate $\kappa_{\mathrm{corr}}$ from a one-dimensional (1D) plasmonic crystal (PlC) connected to a flat metal surface, where the surface plasmon polaritons (SPPs) excited by the incident electron beam propagate over the surface and are converted into photons by the periodic protrusions [6]. Here, we use a silver substrate consisting of one-dimensional grating with 600 nm periodicity with a neighboring flat area, as illustrated in Fig. S3(a). The lattice height of the grating is 50 nm. Figure S3(b) displays the color intensity plot of the CL spectrum as a function of the electron beam position in the line scan. No detection angle selection was performed. The edge position of the grating is set as the origin of the beam position. The PlC features in the spectrum (wavelength axis) are observed in the grating region while only TR is observed in the flat region sufficiently away from the grating. Interferences of SPP and TR appear around the boundary of these two-region, indicating the electron excites both modes [8]. The correlation factor $\kappa_{\mathrm{corr}}$ can be considered as a constant value around one in all the regions including the flat and interfering positions at 200 kV acceleration voltage (Fig. S3(c)). We also observed no acceleration voltage dependence (Fig. S3(d)). Therefore, we can conclude that the photon statistics do not change regardless of the type of coherent radiation from the metal surface, i.e., TR, SPP radiation, or even their mixture with interference where the wavefunction spreads over the surface.

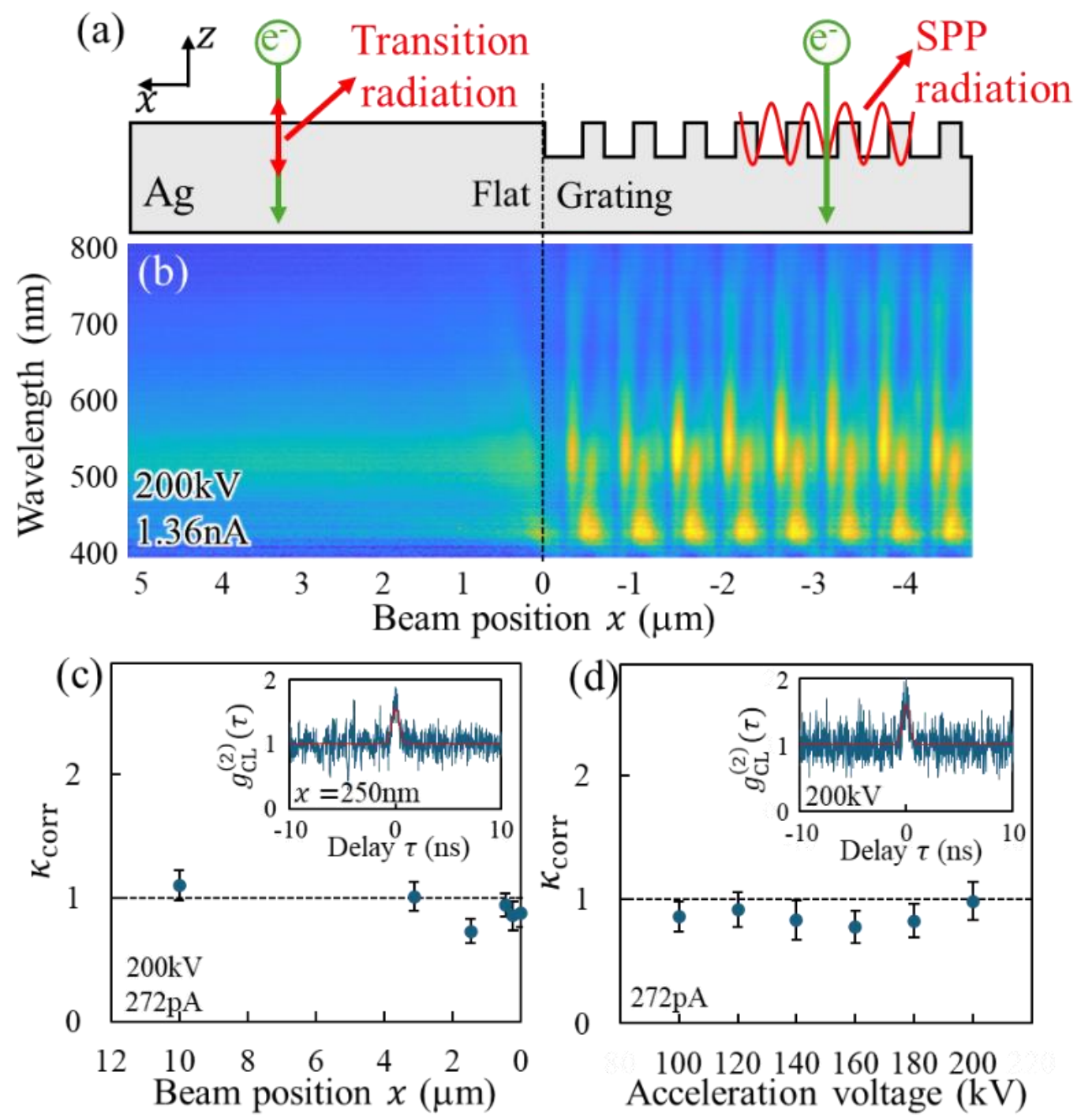

**Fig. S3. Results for a plasmonic crystal structure connected to a flat metal surface.** (a) Schematic illustration of a one-dimensional silver plasmonic crystal (grating) and a flat surface. (b) Color intensity plot of CL spectrum and (c) correlation factor as a function of the beam position. The origin of the beam position (horizontal axis) is set at the boundary of the flat and grating region (see panels (a) and (b)). Inset of (c): correlation curve obtained at a beam position of 250 nm. (d) Acceleration voltage dependency of the correlation factor acquired from the plasmonic crystal region. Inset of (d): correlation curve obtained at the acceleration voltage of 200 kV.

Smith-Purcell Radiation (Coherent CL)

To evaluate the Smith-Purcell (S-P) radiation, we used a one-dimensional periodic structure (periodicity: 800nm, lattice height: 350nm) out of silver, as shown in Fig. S4(a) [9]. The correlation factor $\kappa_{\mathrm{corr}}$ is acquired by scanning the electron beam from the sample area to the vacuum going across the structure as shown in Fig. S4(b). The simultaneously obtained count rate (red line) and the secondary electron signal (green line) profiles are superposed for comparison. The origin of the electron beam position is set at the boundary of the structure and vacuum. We notice that $\kappa_{\mathrm{corr}}$ exceeds one at the position where the electron beam hits the grating structure directly. We attribute this to the reduction of the effective beam current due to the electron beam absorption and scattering by the sample because the focal position of the parabolic mirror is set 10 μm below the sample top surface so that the S-P radiation is efficiently collected. This reduction of the effective beam current increases the apparent value of $\kappa_{\mathrm{corr}}$ as described in Eq. (3) in the main text.

Figures S4(c) shows the correlation factor with different beam currents and acceleration voltages in the aloof excitation with the beam position set at 100 nm from the side edge of the sample. The correlation factors larger than 1 is due to the reduction of the effective beam current because the finite convergence angle of the electron beam and a slight tilt of the sample cause some portion of electrons hitting the sample, resulting in a reduced effective current.

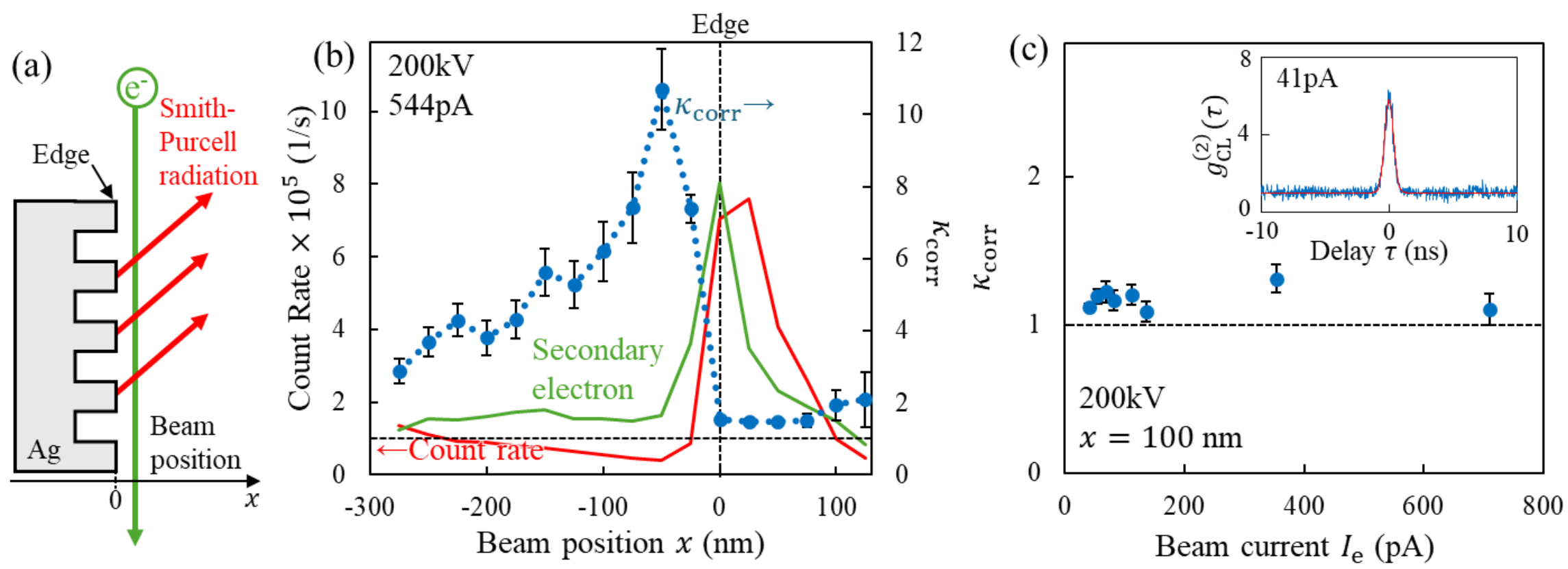

**Fig. S4. Smith-Purcell (S-P) radiation.** (a) Schematic illustration of S-P radiation. The grating has the periodicity of 800 nm and the lattice height of 350nm. (b) Correlation factor plotted as a function of the beam position (blue scattered plot). Secondary electron signal (green line), tracing the geometry of the structure, and photon count rate (red line) profiles are superposed. The edge of the grating is set as the origin of the line scanning measurement. (c) Beam current of the correlation factor obtained in the aloof condition with the electron beam located approximately 100 nm away from the side surface of the sample. The focal position of the parabolic mirror was set 10 μm below the top surface of the sample so that the S-P radiation is efficiently collected. Insets of (c) represent correlation curves obtained at (41 pA, 200 kV).

Cherenkov Radiation (Coherent CL)

The radiation angle $\theta$ of Cherenkov radiation (CR) with respect to the direction of a free electron moving is defined with the refractive index of the medium $n$ and the electron velocity $v$ [10]:

$$\cos\theta = \frac{c}{nv}. \tag{S7}$$

The electron velocity $v$ is a function of acceleration voltage $E$, described as $v = c\sqrt{1-\left(\frac{eE}{m_0c^2}\right)^{-2}}$, where $m_0$ is the rest mass of an electron and $c$ the lightspeed. The threshold electron velocity for CR corresponds to the emission angle $\theta = 0°$. Using a natural mica membrane with the refractive index of ~1.60, the threshold acceleration voltage for CR is calculated as 141 kV [11]. The acceleration voltage dependence of the CR spectrum for mica is shown in Fig. S5. A drastic change in the spectral shape is observed between 140 kV and 200 kV, which is consistent with the CR threshold of 141 kV. Figure S5 suggests that, at 200 kV, the emission from mica is primarily due to CR.

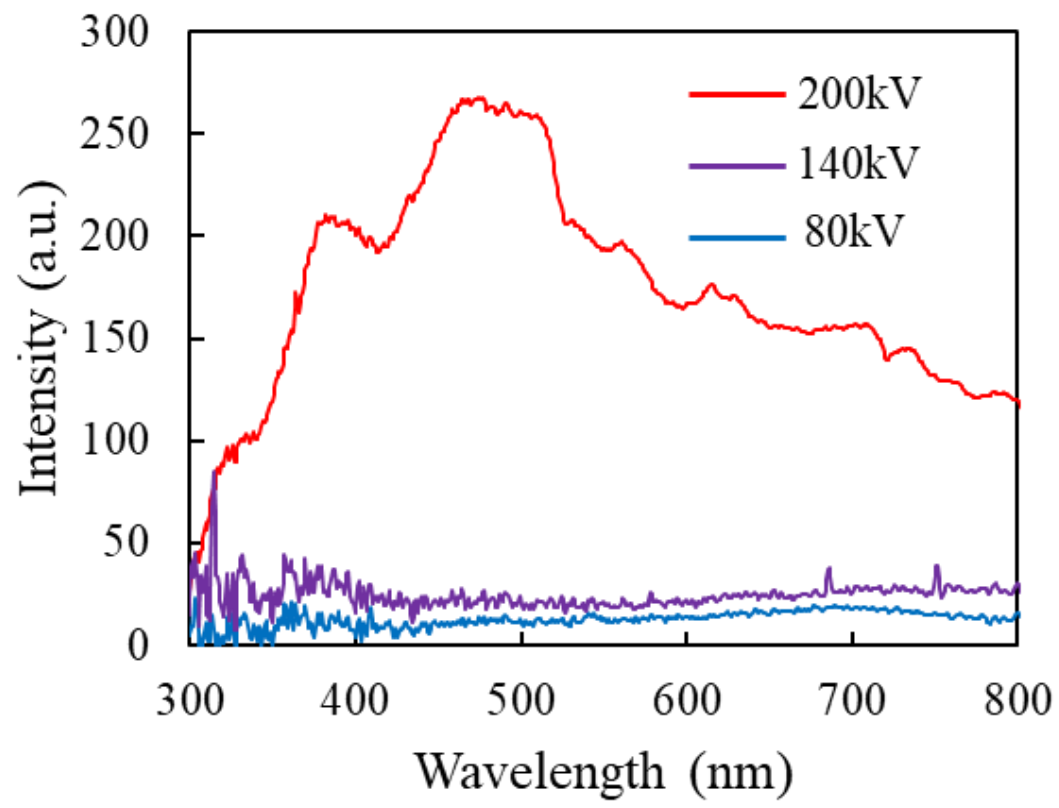

**Fig. S5. CL spectra of Cherenkov radiation from a 140-μm mica sample for different acceleration voltages.** No CL angular selection is applied: the spectra are collected by integrating the signals over the full emission angle.

Thickness of Mica

For the CR measurement, we deposited a 100-200 nm silver layer on the bottom side of the mica membrane and captured the upward-reflected CR. The CR spectra display an etalon pattern that depends on the thickness of mica [12]. By using a pinhole mask to select the radiation angle $\theta$, as shown in Fig. S6(a), the mica thickness $d$ can be calculated as:

$$d = \frac{m}{2\sqrt{n^2 - \sin^2\theta}}\left(\frac{\lambda_1\lambda_2}{\lambda_2 - \lambda_1}\right) \tag{S8}$$

where $\lambda_1$ and $\lambda_2$ are the peak wavelengths of the etalon pattern, $m$ is the number of peaks between the selected $\lambda_1$ and $\lambda_2$, and $n$ is the refractive index of mica. Figure S6 shows the spectra of mica with five different thicknesses mentioned in the main text. The selected emission angle was 53°. As shown in Fig. S6(g), the calculated thickness and the integrated intensity of the spectrum exhibit a proportional relationship, which supports the thickness estimation obtained from the etalon pattern.

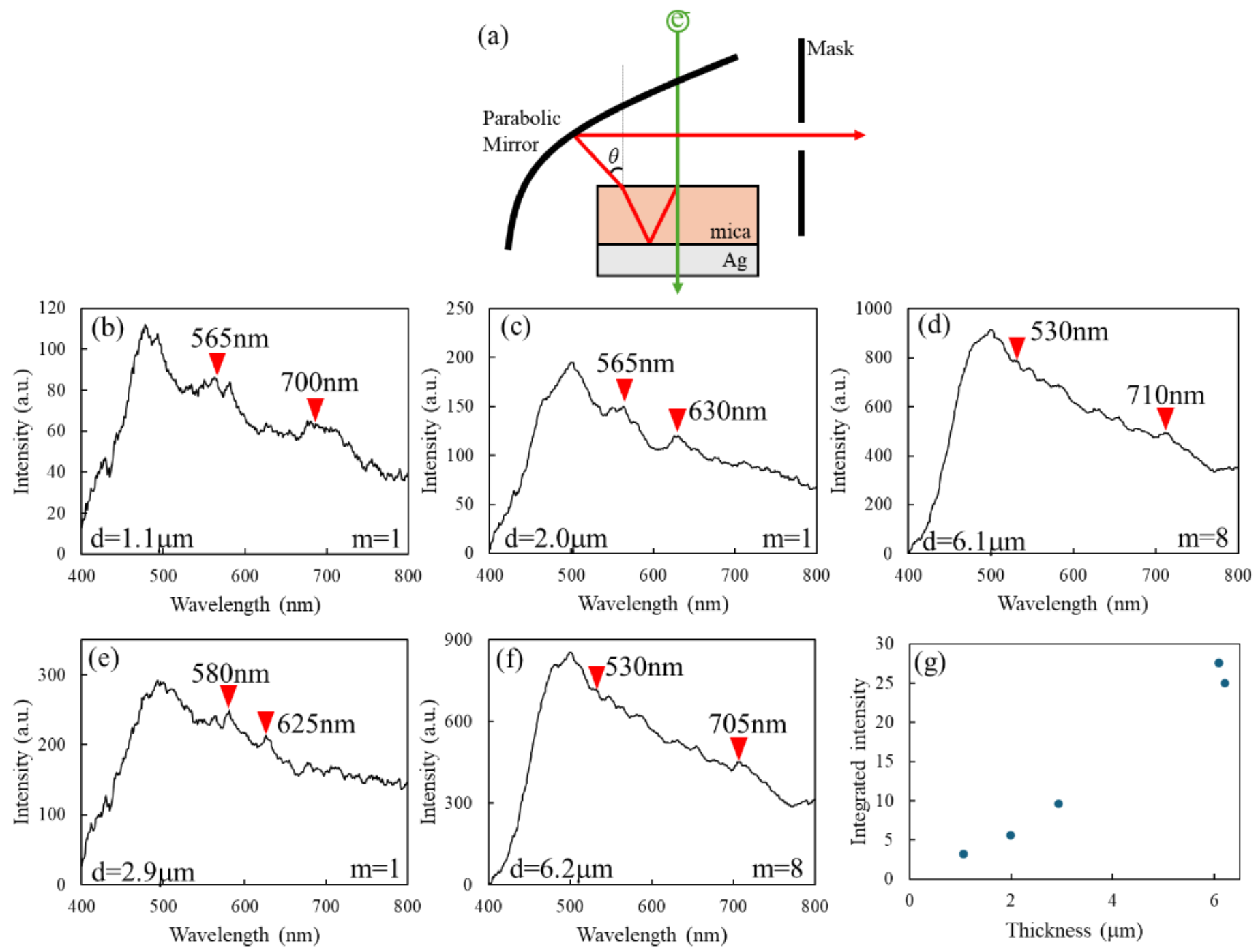

**Fig. S6. Mica spectrum for Cherenkov radiation.** (a) Schematic diagram of the detection setup for CR. (b-f) CL spectra of mica with different thickness obtained at the acceleration voltage of 200kV. The peak wavelengths of the etalon pattern ($\lambda_1$ and $\lambda_2$), the number of peaks between the selected peaks ($m$) and the thickness of the mica ($d$) are indicated. (g) Plot of the integrated intensity of the spectrum (400~800 nm) as the function of the thickness obtained from the etalon pattern.

<u>Comparison of Correlation Function and Correlation Factor for LSP and YSO</u>

Figure S7 shows the results of an electron beam current dependence measurement for coherent CL from LSP, and incoherent CL from YSO particles at an accelerating voltage of 200kV and 80 kV, respectively. We confirmed that $g_{\mathrm{CL}}^{(2)}(0)-1$ is inversely proportional to the electron beam current (with the red solid line representing the fitting), while $\kappa_{\mathrm{corr}}$, extracted from $g_{\mathrm{CL}}^{(2)}(0)$, remains approximately constant.

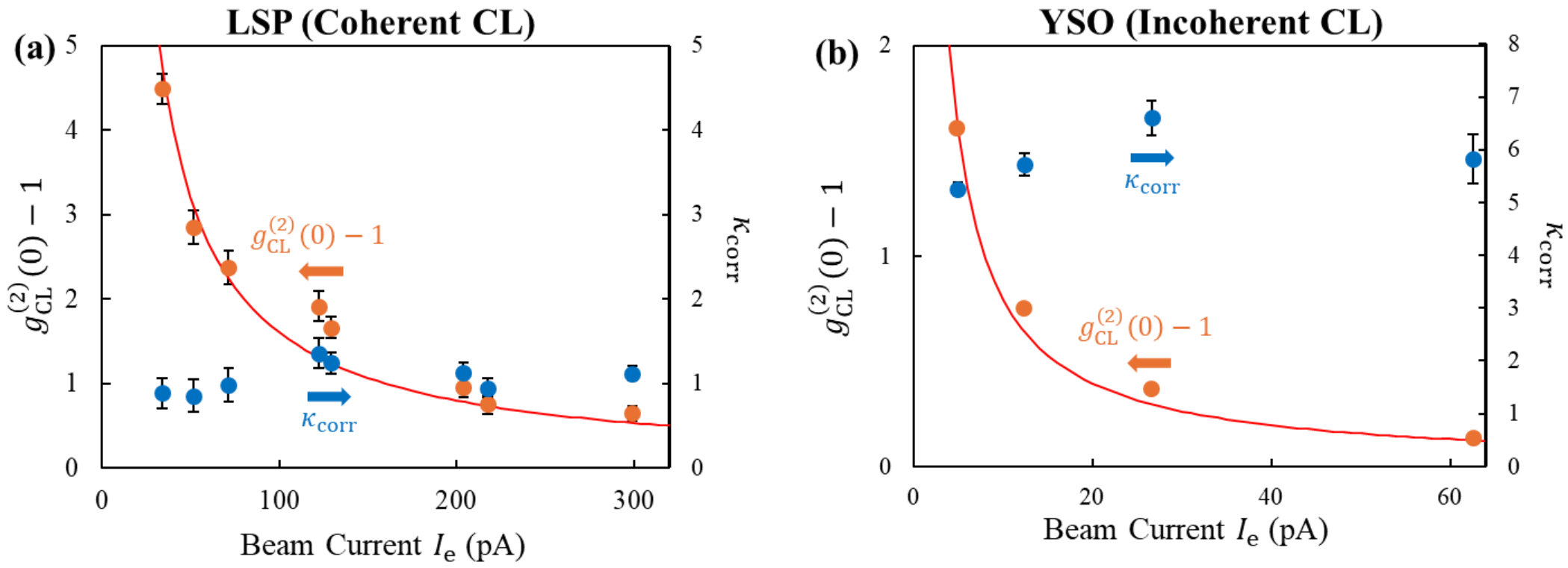

**Fig. S7. Electron beam current dependency of $g_{CL}^{(2)}(0)-1$ and $\kappa_{corr}$ for LSP and YSO.** (a) and (b) show the profiles of $g_{CL}^{(2)}(0)-1$ and $\kappa_{corr}$ as functions of the electron beam current, for LSP (coherent CL) and YSO (incoherent CL), respectively. The red solid line represents the fitting curve based on an inversely proportional function.

Master Equation Approach for Correlation Factor

To derive the correlation factor, incorporating the effect of electron beam current, we here employ a master equation approach. This approach is similar to the chemical master equation, which describes dynamics of chemical reactions [13].

Firstly, we consider the coherent CL with a single excitation process. The "reaction" can be described as follows:

$$\begin{gathered} \emptyset \xrightarrow{\lambda_0} \mathrm{A}_0 \\ \mathrm{A}_0 \xrightarrow{\lambda_{1,m}} m\,\mathrm{A}_1 \\ \mathrm{A}_1 \xrightarrow{\gamma_1} \emptyset \end{gathered} \quad . \tag{S9}$$

$\mathrm{A}_0$ is an electron incident into the "reaction chamber" with the rate of $\lambda_0$ [s$^{-1}$], corresponding to the beam current ($\lambda_0 = I_e/e$). The single electron ($\mathrm{A}_0$ particle) generates $m$ photons ($\mathrm{A}_1$ particles) with the rate of $\lambda_{1,m}$ [s$^{-1}$], where $m$ varies stochastically. Each of the generated photons disappears with the rate of $\gamma_1$ [s$^{-1}$], which corresponds to the decay time ($\gamma_1 = 1/\tau_w$). This leads to a master equation:

$$\frac{dP(n_0,n_1;t)}{dt} = \lambda_0 P(n_0-1,n_1;t) - \lambda_0 P(n_0,n_1;t) + \sum_{m=0}^{n_1}\{\lambda_{1,m}(n_0+1)P(n_0+1,n_1-m_1;t)\} - \lambda_1 n_0 P(n_0,n_1;t)$$

$$+\gamma_1(n_1+1)P(n_0,n_1+1;t)-\gamma_1 n_1 P(n_0,n_1;t) \quad . \tag{S10}$$

$P(n_0,n_1;t)$ is a probability of having $n_0$ number of $\mathrm{A}_0$ particles (electrons) and $n_1$ number of $\mathrm{A}_1$ particles (photons) at time $t$. We also define $\lambda_1=\sum_{m=0}^{\infty}\lambda_{1,m}$. The time derivative of the expectation value of $n_0$, $\langle n_0\rangle=\sum_{n_0,n_1} n_0 P(n_0,n_1;t)$, should equal to zero for the steady state, which is denoted by suffix "ss":

$$\begin{aligned}\frac{d\langle n_0\rangle}{dt}&=\lambda_0\langle n_0+1\rangle-\lambda_0\langle n_0\rangle+\lambda_1\langle(n_0-1)n_0\rangle-\lambda_0\langle {n_0}^2\rangle+\gamma_1\langle n_0n_1\rangle-\gamma_1\langle n_0n_1\rangle\\ &=\lambda_0-\lambda_1\langle n_0\rangle \quad \xrightarrow{\text{steady state}} \langle n_0\rangle_{\text{ss}}=\lambda_0/\lambda_1 \quad .\end{aligned} \tag{S11}$$

In the same manner, we obtain the following expressions:

$$\begin{gathered}\langle n_1\rangle_{\text{ss}}=\langle m_1\rangle_{1\mathrm{A}_0}\lambda_0/\gamma_1 \ , \\ \langle {n_0}^2\rangle_{\text{ss}}=\left(\frac{\lambda_0}{\lambda_1}\right)^2+\frac{\lambda_0}{\lambda_1} \ , \\ \langle n_0n_1\rangle_{\text{ss}}=\frac{1}{\lambda_1+\gamma_1}\left\{\lambda_1\langle m_1\rangle_{1\mathrm{A}_0}\left(\langle {n_0}^2\rangle_{\text{ss}}-\langle n_0\rangle_{\text{ss}}\right)+\lambda_0\langle n_1\rangle_{\text{ss}}\right\} \ , \\ \langle {n_1}^2\rangle_{\text{ss}}=\frac{1}{2\gamma_1}\left\{\lambda_1\langle {m_1}^2\rangle_{1\mathrm{A}_0}\langle n_0\rangle_{\text{ss}}+\gamma_1\langle n_1\rangle_{\text{ss}}+2\lambda_1\langle m_1\rangle_{1\mathrm{A}_0}\langle n_0n_1\rangle_{\text{ss}}\right\} \ ,\end{gathered} \tag{S12}$$

where $\langle m_1\rangle_{1\mathrm{A}_0}=\sum_m m\frac{\lambda_{1,m}}{\lambda_1}$ corresponds to the average number of generated $\mathrm{A}_1$ (photon) particles per $\mathrm{A}_0$ particle (electron) and $\langle {m_1}^2\rangle_{1\mathrm{A}_0}=\sum_m m^2\frac{\lambda_{1,m}}{\lambda_1}$. Using the expressions above, we obtain

$$g_{\text{CL}}^{(2)}(0)=\frac{\langle {n_1}^2\rangle_{\text{ss}}-\langle n_1\rangle_{ss}}{{\langle n_1\rangle_{\text{ss}}}^2}=1+\frac{\gamma_1}{2\lambda_0}\frac{\langle {m_1}^2\rangle_{1\mathrm{A}_0}-\langle m_1\rangle_{1\mathrm{A}_0}}{{\langle m_1\rangle_{1\mathrm{A}_0}}^2}=1+\frac{e}{2\tau_w I_e}\kappa_{\text{corr}} \ . \tag{S13}$$

Eq. S13 is identical to Eq. 3 in the main text.

Secondly, we consider the incoherent CL with a two-step excitation process. For this process the "reaction" is described as,

$$\begin{gathered}\emptyset \xrightarrow{\lambda_0} \mathrm{A}_0 \\ \mathrm{A}_0 \xrightarrow{\lambda_{1,m_1}} m_1\mathrm{A}_1 \\ \mathrm{A}_1 \xrightarrow{\lambda_{2,m_2}} m_2\mathrm{A}_2 \\ \mathrm{A}_2 \xrightarrow{\gamma_2} \emptyset\end{gathered} \quad . \tag{S14}$$

Here $\mathrm{A}_0$ is again the incident electron, $\mathrm{A}_1$ is now the secondary (mediator) particle, and $\mathrm{A}_2$ is the photon. Repeating almost the same calculation as that for the coherent CL, we can derive the following expression:

$g^{(2)}(0) =$

$$1+\frac{\lambda_2\gamma_2}{2\lambda_0(\lambda_2+\gamma_2)}\left\{\frac{\langle {m_1}^2\rangle_{1\mathrm{A}_0}-\langle m_1\rangle_{1\mathrm{A}_0}}{{\langle m_1\rangle_{1\mathrm{A}_0}}^2}+\frac{\lambda_2+\gamma_2}{\lambda_2}\frac{1}{\langle m_1\rangle_{1\mathrm{A}_0}}\frac{\langle {m_2}^2\rangle_{1\mathrm{A}_1}-\langle m_2\rangle_{1\mathrm{A}_1}}{{\langle m_2\rangle_{1\mathrm{A}_1}}^2}\right\}, \quad \text{(S15)}$$

where $\langle m_k\rangle_{1\mathrm{A}_{k-1}} = \sum_{m_k} m_k \frac{\lambda_{k,m_k}}{\lambda_k}$ and $\langle {m_k}^2\rangle_{1\mathrm{A}_{k-1}} = \sum_{m_k} {m_k}^2 \frac{\lambda_{k,m_k}}{\lambda_k}$ with $\lambda_k = \sum_{m_k=0}^{\infty} \lambda_{k,m_k}$ ($k = 1, 2$). We can reasonably assume that the "reaction" of generating $\mathrm{A}_2$ (photons) by secondary $\mathrm{A}_1$ particles is much faster than the decay of $\mathrm{A}_2$ (photons), i.e., $\gamma_2 \ll \lambda_2$. This leads to

$$g_{\mathrm{CL}}^{(2)}(0) = 1+\frac{e}{2\tau_w I_e}\left\{\kappa_{\mathrm{corr}}^{(1)}+\frac{1}{\langle m_1\rangle_{1\mathrm{A}_0}}\kappa_{\mathrm{corr}}^{(2)}\right\}, \quad \text{(S16)}$$

which is identical to the expression of Eq. 3 in the main text. We can derive the general expression of Eq. 6 for the multi-step excitation process in the same manner, also considering the multi-step treatment of Eq. S5.

**References**

1. Polman, A., Kociak, M. & García de Abajo, F. J. Electron-beam spectroscopy for nanophotonics. *Nat Mater* **18**, 1158–1171 (2019).
2. Kfir, O. Entanglements of Electrons and Cavity Photons in the Strong-Coupling Regime. *Phys Rev Lett* **123**, 103602 (2019).
3. García de Abajo, F. J. Multiple excitation of confined graphene plasmons by single free electrons. *ACS Nano* **7**, 11409–11419 (2013).
4. Loudon, R. *The Quantum Theory of Light*. (OUP Oxford, 2000).
5. García De Abajo, F. J. Optical excitations in electron microscopy. *Rev Mod Phys* **82**, 209–275 (2010).
6. Yamamoto, N. Development of high-resolution cathodoluminescence system for STEM and application to plasmonic nanostructures. *Microscopy* **65**, 282–295 (2016).
7. Palik, E. D. *Handbook of Optical Constants of Solids*. (Academic press, 1985).
8. Takeuchi, K. & Yamamoto, N. Visualization of surface plasmon polariton waves in two-dimensional plasmonic crystal by cathodoluminescence. *Opt Express* **19**, 12365 (2011).
9. Yamamoto, N., García de Abajo & Myroshnychenko, V. Interference of surface plasmons and Smith-Purcell emission probed by angle-resolved cathodoluminescence spectroscopy. *Phys Rev B* **91**, 125144 (2015).
10. Yamamoto, N., Sugiyama, H. & Toda, A. Cherenkov and transition radiation from thin plate crystals detected in the transmission electron microscope. *Proceedings of the Royal Society A: Mathematical, Physical and Engineering Sciences* **452**, 2279–2301 (1996).
11. El-Zaiat, S. Y. Application of multiple-beam white-light fringes for measuring the refraction and dispersion of mica. *Opt Laser Technol* **29**, 495–500 (1998).
12. Bashevoy, M. V. *et al*. Generation of traveling surface plasmon waves by free-electron impact. *Nano Lett* **6**, 1113–1115 (2006).
13. Van Kampen, N. G. *Stochastic Processes in Physics and Chemistry*. (Elsevier, 1981).